\documentclass[aps,10pt,superscriptaddress,amsmath,amssymb,twocolumn,showpacs,floatfix,reprint]{revtex4-2}
\usepackage{braket}
\usepackage{subfigure}
\usepackage[latin9]{inputenc}
\usepackage{color}
\usepackage{verbatim}
\usepackage{amsthm,amsmath,amssymb}
\usepackage{mathrsfs}
\usepackage{graphicx}
\usepackage{esint}
\usepackage{mathrsfs}
\usepackage{bm}
\usepackage{hyperref}
\usepackage{mathrsfs}

\begin{document}
\title{Chiral spin liquids on the kagome lattice with projected entangled simplex states}

\author{Sen Niu}\email{sen.niu@irsamc.ups-tlse.fr}
\affiliation{Laboratoire de Physique Th\'eorique, C.N.R.S. and Universit\'e de Toulouse, 31062 Toulouse, France}

\author{Juraj Hasik}
\affiliation{Institute for Theoretical Physics, University of Amsterdam,
Science Park 904, 1098 XH Amsterdam, The Netherlands}

\author{Ji-Yao Chen}\email{chenjiy3@mail.sysu.edu.cn}
\affiliation{School of Physics, Sun Yat-sen University, Guangzhou, 510275, China}
\affiliation{Dahlem Center for Complex Quantum Systems, Freie Universit\"at Berlin, Berlin 14195, Germany}

\author{Didier Poilblanc}
\affiliation{Laboratoire de Physique Th\'eorique, C.N.R.S. and Universit\'e de Toulouse, 31062 Toulouse, France}

\begin{abstract}
The infinite projected entangled simplex state (iPESS), a type of tensor network (TN) state, has been used successfully for simulating and characterizing {\it non-chiral} spin liquids on the kagome lattice. Here, we demonstrate that iPESS also provides a faithful representation of a {\it chiral} spin liquid (CSL) on the same lattice, namely the ground state of the spin-$1/2$ kagome Heisenberg antiferromagnet with a scalar chirality. By classifying local tensors according to SU$(2)$ and point group symmetries, we construct a chiral ansatz breaking reflection $P$ and time reversal $T$ symmetries while preserving $PT$. The variational TN states are shown to host, for bond dimension $D\ge 8$, a chiral gapless entanglement spectrum following SU$(2)_1$ conformal field theory. The correlation function shows a small weight long-range tail complying with the prediction of the TN bulk-edge correspondence. 
We identify a non-chiral manifold spanned by only a subset of symmetric tensors where a new emergent {\it tensor conservation law} is realized. This allows us to both probe the stability of the non-chiral spin liquid and discuss its transition to CSL induced by a scalar chirality term.

\end{abstract}
\maketitle

\emph{Introduction.---}Quantum spin liquids have received numerous attention in condensed matter theory and experiments due to their exotic properties such as fractionalized excitations and long-range entanglement~\cite{savary2016quantum,zhou2017quantum}. A prominent system is the kagome Heisenberg antiferromagnet (KHA) where the quantum ground state is non-magnetic. Promising ground states were proposed such as the algebraic U$(1)$ spin liquid with Dirac spinon spectrum~\cite{ran2007projected} and the gapped spin liquid with $\mathbb{Z}_2$ topological order~\cite{sachdev1992kagome}. By  perturbing such system with a scalar spin chirality term or longer-range Heisenberg couplings, a more intriguing chiral spin liquid (CSL) state can be stabilized as uncovered by density-matrix renormalization group studies in quasi-one dimension~\cite{kagomeCSL2014,gong2014emergent}.
CSL states were proposed by Kalmeyer and Laughlin in spin models as a bosonic variant of some fractional Quantum Hall state~\cite{tsui1982two}. Such states host anyonic quasiparticles~\cite{halperin1984statistics} in the bulk as well as chiral gapless modes~\cite{wen1991gapless} on the edge described by ($1 + 1$)-dimensional conformal field theories (CFT).

In recent decades, tensor network (TN) states ~\cite{verstraete2008matrix} have emerged as a  powerful framework for investigating strongly correlated systems both analytically and numerically. 
They encode many-body state into a network of local tensors contracted by their connected virtual bonds. 
In two dimensions, TN states can represent several well-known \emph{non-chiral} spin liquid states exactly, e.g., the toric code state with $\mathbb{Z}_2$ topological order~\cite{toriccode2006,toriccode2009}, the string-net states~\cite{stringnetnet2009}, the short-range resonating valence bond (RVB) states on the kagome lattice (with $\mathbb{Z}_2$ topological order) and on the square lattice (being gapless and with U$(1)$ gauge symmetry)~\cite{poilblanc2012topological}. On the numerical side, TN states have also proven their competitiveness against other numerical methods~\cite{iqbal2013} in determining the nature of ground states of challenging models such as the KHA~\cite{KHA2017gap,KHA2017gapless}. 

In contrast to the case of non-chiral topological states where TN representation works well with local gauge symmetry imposed~\cite{toriccode2009,formalism2013}, it is less clear whether TN states can represent generic chiral topological states~\cite{wahl2013projected,yang2015chiral,dubail2015tensor}. In the past, a lot of effort has been devoted to construct CSL states on the square lattice~\cite{poilblanc2015chiral,systematic2016,poilblanc2017,chen2020,chen2021abelian,hasik2022,chen2018non} with symmetric ~\cite{Jiang2015,systematic2016} infinite projected entangled pair states (iPEPS)~\cite{verstraete2004renormalization,jordan2008classical}. These constructions include two important ingredients. First, both the reflection $P$ and time reversal $T$ symmetries are broken while the combined $PT$ symmetry is preserved. On the square lattice with $C_{4v}$ point group symmetry, one can choose the real elementary tensors transforming according to $\mathrm{A}_1$ and $\mathrm{A}_2$ irreducible representations (IRREPs)~\cite{systematic2016}, and the $\mathrm{A}_1+i\mathrm{A}_2$ chiral ansatz  then satisfies the symmetry condition exactly. Second, appropriate spins should be assigned in the virtual space. For the simplest SU$(2)_1$ CSL (bosonic $\nu=1/2$ Laughlin state), the virtual space is chosen as a direct sum $\mathcal{V}=0\oplus 1/2$ where two virtual spins $0$ and $1/2$ correspond to two primary fields of Wess-Zumino-Witten (WZW) SU$(2)_1$ CFT and the local tensor satisfies the necessary gauge symmetry~\cite{poilblanc2015chiral}. This heuristic rule for virtual spins turns out to be valid also for SU$(N)_1$~\cite{chen2021abelian} and non-Abelian SU$(2)_2$~\cite{chen2018non} CSLs. 

However, generalizing the construction to the non-bipartite kagome lattice is not straightforward. The natural extension of iPEPS on the kagome lattice~\cite{schuch2012} was named infinite projected entangled simplex state (iPESS)~\cite{xie2014tensor}. A direct adoption of the heuristic rule fails for the simplest SU$(2)_1$ case as the trivalent tensor ansatz $\mathrm{A}_1+i\mathrm{A}_2$ with virtual space $\mathcal{V}=0\oplus 1/2$ yields the non-chiral short-range RVB state~\cite{poilblanc2012topological}. Here the phase $i$ only contributes a global phase to the many-body state due to an emergent conservation law for the total number of $\mathrm{A}_1$ tensors on the lattice. One then has to generalize the heuristic rule by including more relevant virtual spins while keeping the gauge symmetry intact in the local tensors. 
The main purpose of this paper is to provide the first example of a two-dimensional TN (iPESS) construction of CSL states on the kagome lattice. We start from the Hamiltonian which hosts a CSL ground state.

\emph{Model and classification of symmetric iPESS.---}We consider the nearest neighbour spin-$1/2$ KHA model with a scalar spin chirality term $\chi_{ijk}=\mathbf{S}_i\cdot (\mathbf{S}_j\times\mathbf{S}_k)$ acting on both up and down triangles 
\begin{align}
H= J_H\sum_{\langle i,j\rangle}\mathbf{S}_i \cdot \mathbf{S}_j+J_{\chi}\sum_{i,j,k\in\bigtriangleup,\bigtriangledown} \chi_{ijk},
\label{eq:model}
\end{align}
with sites $i,j,k$ on triangles ordered clockwise. The $\chi_{ijk}$ term breaks $P$ and $T$ symmetries, while preserving SU$(2)$ spin rotation and $PT$ symmetries. For convenience, we define $J_{\chi}/J_H=\tan{\theta}$ and set $J_H=1$. The $\theta=0$ point then corresponds to the $T$-symmetric KHA point. 

We construct the iPESS ansatz for a translationally invariant state on the kagome lattice from two unique  tensors: An on-site tensor $b$ which resides on physical sites and a trivalent tensor $t$ which resides in triangles. Using Penrose notation, $b$ and $t$ have the form
\begin{equation}
\includegraphics[scale = 0.7]{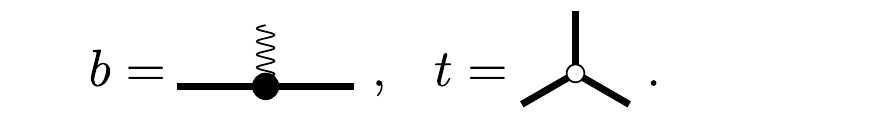} \label{eq:diagram0}
\end{equation}
Here, the on-site $b$ tensor
has one physical index (wavy line) of dimension $2$ corresponding to spin-$1/2$ and two virtual indices (solid lines) of dimension $D$ running over states in the virtual space $\mathcal{V}$, while the trivalent tensor $t$
has three virtual indices of dimension $D$. Kagome lattice is tiled with $t$ tensors, each connected by  three $b$ tensors to neighbouring $t$'s as shown below.  For evaluating the physical observables, we group three adjacent $b$ tensors and two $t$ tensors into an iPEPS tensor $a$ defined as
\begin{equation}
\includegraphics[scale = 0.7]{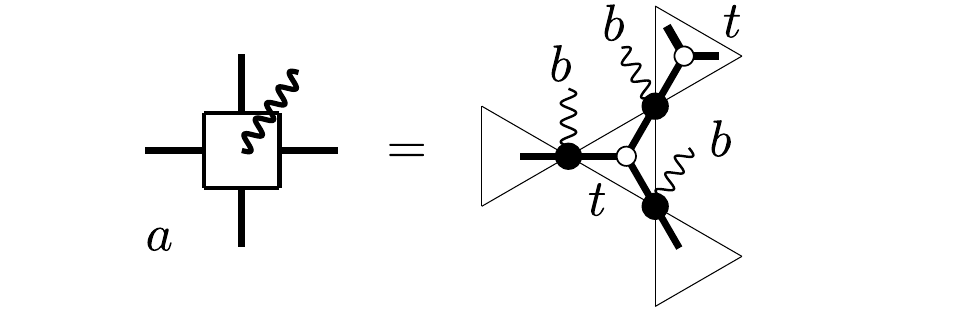}, \label{eq:ipepstensor}
\end{equation}
mapping the wave function to an effective square lattice where each unit-cell contains three spins. We adopt the corner transfer matrix renormalization (CTMRG) method~\cite{ctmrg1,ctmrg2}, controlled by environment dimension $\chi$, to contract the resulting TN approximately (exact in the $\chi\rightarrow\infty$ limit). 

\begin{table}[t]
\begin{tabular}{|c|c|c|c|c|c|c|c|c|}
\hline
${D}^*$&$D$ & Virtual space & $b_{\rm A}$ & $b_{\rm B}$ & $t_{\rm A_{1}}^{\text{I}}$ & $t_{\rm A_{1}}^{\text{II}}$ & $\rm t_{A_{2}}^{\text{I}}$ & $t_{\rm A_{2}}^{\text{II}}$ \tabularnewline
\hline
2&3 & $0\oplus\frac{1}{2}$ & 1 & 1 & 1 &0 & 0 &1 \tabularnewline
\hline
3&6 & $0\oplus\frac{1}{2}\oplus1$ & 2 & 2 & 1 &2 & 1 &1 \tabularnewline
\hline
4&8 & $0\oplus\frac{1}{2}\oplus 1 \oplus \frac{1}{2}$ & 4 & 4 & 2 & 4 & 1 &4 \tabularnewline
\hline
5&12 & $0\oplus\frac{1}{2}\oplus 1 \oplus \frac{1}{2}\oplus\frac{3}{2}$ & 5 & 5 & 2 &7 & 1 &7 \tabularnewline
\hline
\end{tabular}
\caption{Virtual spaces $\mathcal{V}$ for different multiplet dimensions ${D}^*$ and numbers of distinct elementary tensors resolved by point group symmetry (see text).}
\label{table:tensors}
\end{table}

To further restrict the ansatz, we impose both point group symmetries and SU$(2)$ symmetry on both $b$ and $t$ tensors ensuring that the many-body state preserves lattice symmetries and is a global spin singlet.
The latter requires a choice of an appropriate virtual space $\mathcal{V}$ for the CSL state, which we obtain from  simple update simulations~\cite{Jiang2008,KHA2017gap} at $\theta=0.2\pi$, deep in the CSL  phase~\cite{kagomeCSL2014}. Starting from a $D=6$ ansatz with $\mathcal{V}=0\oplus \frac{1}{2} \oplus 1$ ~\cite{KHA2017gap,Jiang2019} and appropriate point group quantum number (see below) we perform the imaginary time evolution while keeping SU$(2)$ symmetry.
The resulting optimal virtual spaces $\mathcal{V}=\oplus_{i=1}^{D^*}\mathcal{V}_i$ for various multiplet dimension $D^*$ are shown in table~\ref{table:tensors}.
The product virtual space $\mathcal{V}^{\otimes 2}$ for $b$ ($\mathcal{V}^{\otimes 3}$ for $t$) tensor can be decomposed into disconnected subspaces labeled by the occupations $\{n_1, . . . ,n_{D^*}\}$ 
with the constraint $\sum_{i}n_i=2$ ($\sum_{i}n_i=3$). Importantly, due to SU$(2)$ invariance the total occupations for even (integer) virtual spins $n_{\rm even}$ and odd (half-integer) virtual spins $n_{\rm odd}$ are constrained. The $b$ tensors satisfy $\{n_{\rm even},n_{\rm odd}\}=\{1,1\}$ due to the physical spin $S=1/2$. For the $t$ tensor, $\{n_{\rm even},n_{\rm odd}\}$ can be either $\{3,0\}$ (dubbed as $t^\mathrm{I}$) or $\{1,2\}$ (dubbed as $t^\mathrm{II}$). 
Graphically these tensors take the form
\begin{equation}
\includegraphics[scale = 0.7]{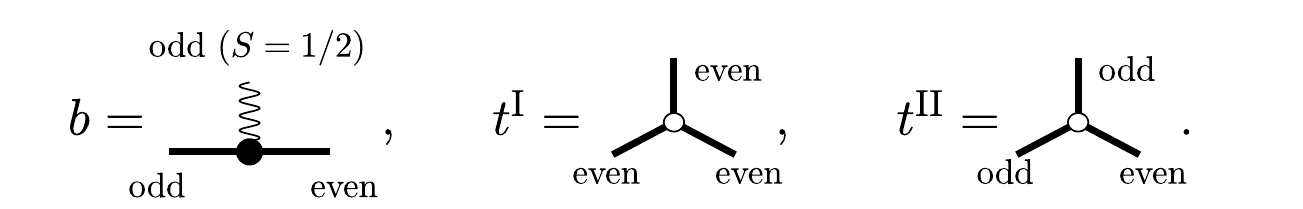} \label{eq:diagram_combined}
\end{equation}
The conservation of the total {\it parity} for $b$ and $t$ tensors infers a $\mathbb{Z}_2$ gauge symmetry to our family of TN states.

Finally, we impose point group symmetries $C_2$ on $b$ and $C_{3v}$ on $t$ tensors  w.r.t permutation of virtual indices. The $b$ tensors are divided into $b_{\mathrm{A}}$ and $b_{\mathrm{B}}$, transforming as symmetric $\mathrm{A}$ and anti-symmetric $\mathrm{B}$ IRREPs of the $C_{2}$ group. Similarly, $t$ tensors are further divided into $t^{\mathrm{I}}_{\mathrm{A_1}},t^{\mathrm{I}}_{\mathrm{A_2}},t^{\mathrm{II}}_{\mathrm{A_1}},t^{\mathrm{II}}_{\mathrm{A_2}}$, where $\mathrm{A}_1,\mathrm{A}_2$ are reflection symmetric and anti-symmetric IRREPs of the $C_{3v}$ group.
The complete set of these real-valued elementary tensors can be obtained from the diagonalization procedure proposed in Ref.~\cite{systematic2016}. We show the number of these elementary tensors in each subspace in Tab.~\ref{table:tensors}.

Now we are ready to construct the many-body chiral/non-chiral states by defining $b$ and $t$ tensor as superposition of elementary tensors. 
For $b$ we consider only $b_{\mathrm{A}}$ elementary tensors, since they are equivalent to  $b_{\mathrm{B}}$ via a virtual $\mathbb{Z}_2$ gauge transformation~\cite{hackenbroich2018interplay} and their superposition breaks lattice rotation symmetry. The existence of reflection symmetry then depends only on the choice of the $t$ tensors. There are two types of non-chiral states constructed from (i) pure $\mathrm{A}_{1}$ or pure $\mathrm{A}_{2}$ IRREP tensors or (ii) $t^{\mathrm{I}}_{\mathrm{A}_{1}}+e^{i\phi}t^{\mathrm{II}}_{\mathrm{A}_{2}}$ or $t^{\mathrm{II}}_{\mathrm{A}_{1}}+e^{i\phi}t^{\mathrm{I}}_{\mathrm{A}_{2}}$ tensors, $\phi$ being an (irrelevant) arbitrary number. The non-chiral ansatz (ii) preserves reflection symmetry due to an emergent \emph{conservation law} for the number of $\mathrm{A}_1$ tensors on the lattice: Every $b$ tensor has one odd virtual index and, on the lattice with a fixed number of physical sites, the number of odd virtual indices is also fixed. Therefore, the number of $t^{\mathrm{II}}$ tensors is conserved and the $e^{i\phi}$ factor only changes the global phase of the many-body state. The ansatz for chiral states should allow for fluctuations of the number of $t^{\mathrm{I}}_{\mathrm{A}_{1}}$ and $t^{\mathrm{II}}_{\mathrm{A}_{2}}$ tensors, which is realized by choosing $t^{\mathrm{I}}_{\mathrm{A}_{1}}+t^{\mathrm{II}}_{\mathrm{A}_{1}}+it^{\mathrm{I}}_{\mathrm{A}_{2}} +it^{\mathrm{II}}_{\mathrm{A}_{2}}$ (dubbed as $\mathrm{A}_1+i\mathrm{A}_2$) which is invariant under $PT$ operation.

\begin{figure}[t]
\centering
\includegraphics[width=\columnwidth]{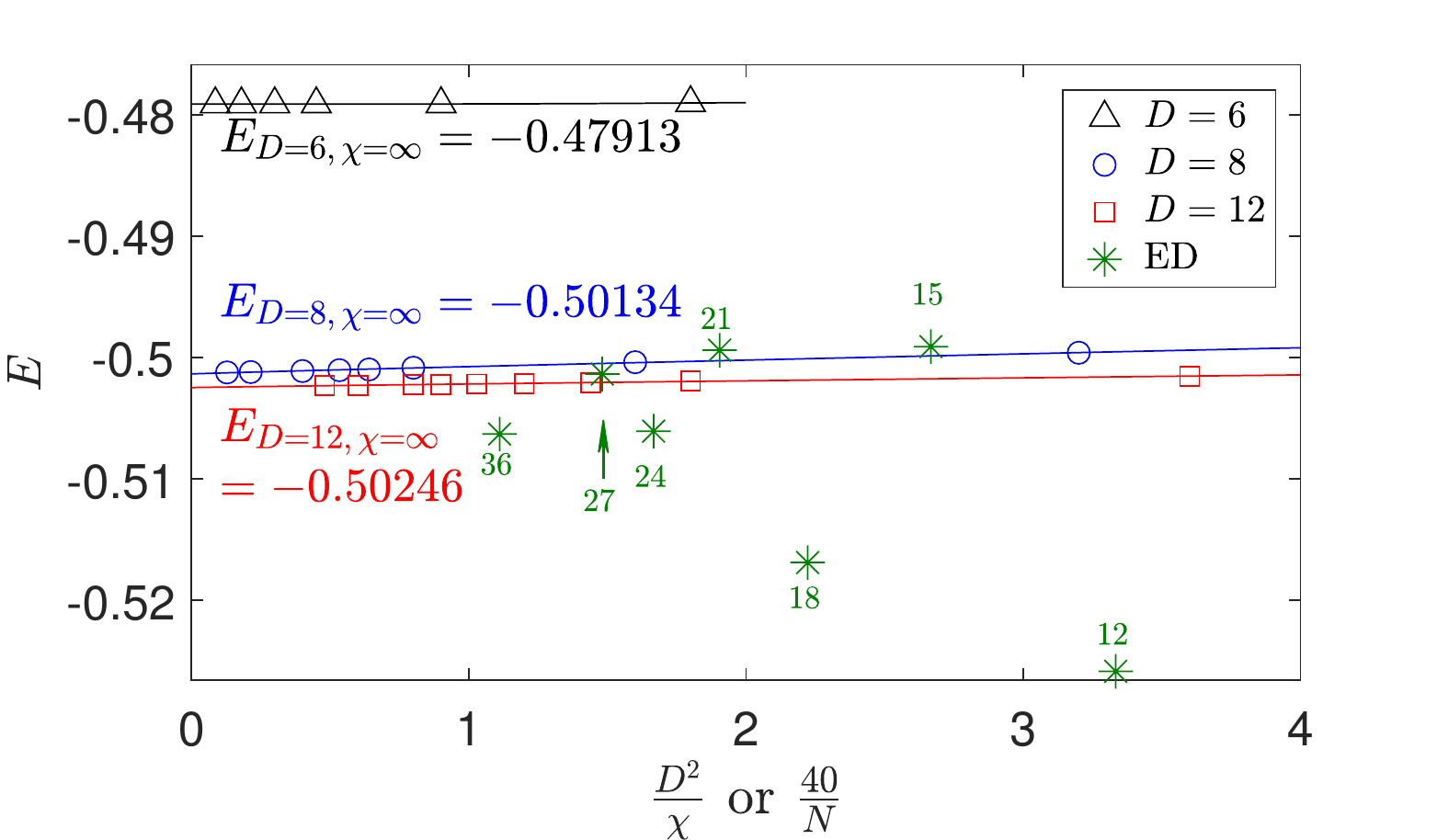}
\caption{Energies of the $\mathrm{A}_1+i\mathrm{A}_2$ chiral iPESS at $\theta=0.2\pi$ as functions of CTMRG environment dimension $\chi$. The states are optimized at $\chi=40,60,120$ for $D=6,8,12$. The ED energies on clusters with sizes ($N=12,15,18,21,24,27,36$) are marked by green symbols.}
\label{fig:energy_0d2pi}
\end{figure}

\emph{CSL state in the iPESS representation.---} Here we demonstrate that the topological CSL state can be represented by the $\mathrm{A}_1+i\mathrm{A}_2$ chiral iPESS by determining it variationally at  $\theta=0.2\pi$, deep in the CSL phase. In our optimization scheme, the $b$ and $t$ tensors are written as linear superpositions with real coefficients of the elementary tensors of Tab.~\ref{table:tensors}. These coefficients are the variational parameters and we optimize them by gradient descent method, where the gradients are obtained from finite differences of energy evaluated by CTMRG at fixed $\chi$. The energies at $D=6,8,12$ in Fig.~\ref{fig:energy_0d2pi} show that the ground state energy converges quickly. By comparing with the exact diagonalization (ED) energies on finite-size clusters (up to $36$ sites) we find that already from $D\geq8$ the chiral iPESS ansatz has very good energy.

\begin{figure}[t]
\centering
\includegraphics[width=\columnwidth]{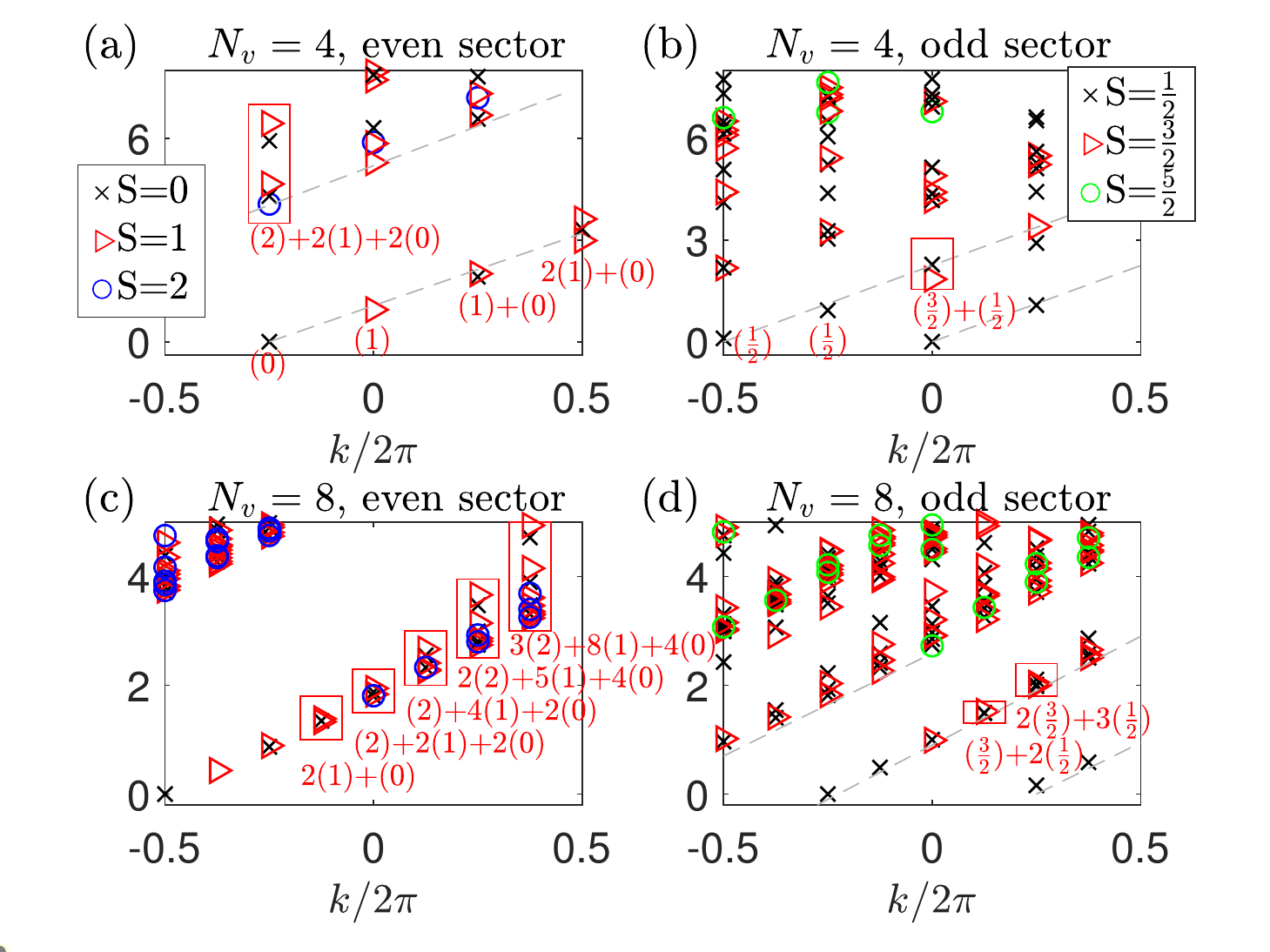}
\caption{Entanglement spectrum of the optimized $\mathrm{A}_1+i\mathrm{A}_2$ chiral iPESS ($D=8$) at $\theta=0.2\pi$ with even/odd sectors normalized separately. (a, b) ES on width $N_v=4$ cylinders obtained by ED. (c, d) ES on $N_v=8$ cylinders  obtained from CTMRG with $\chi =40$. The levels marked by red boxes agree with the level counting of the SU$(2)_1$ WZW CFT.}
\label{fig:ES}
\end{figure}

In order to identify the topological nature of optimized states, we compute the bipartite entanglement spectrum (ES) as proposed by Li and Haldane~\cite{li2008entanglement}. 
Here, the iPEPS tensors $a$ of Eq.~\ref{eq:ipepstensor} are put on an infinitely long cylinder with finite circumference $N_v$. The entanglement spectrum of the TN state can then be calculated exactly by ED for small $N_v$~\cite{cirac2011entanglement,poilblanc2012topological} and approximately with the CTMRG method~\cite{poilblanc20162,chen2018non,chen2020} for larger $N_v$.
We find that a satisfactory agreement is reached  between the ES of $D\ge 8$ states and the predictions of SU$(2)_1$ CSL~\footnote{Note the ES computed at $D=6$ does not show the CSL features, lacking well separated linear branches.}. The results of ES for the $D=8$ variational state are displayed in Fig.~\ref{fig:ES}. The levels are labeled by momentum and total spin quantum numbers. One can see the low-lying chiral branches disperse linearly and the levels in the even/odd sector match those of two SU$(2)_1$ conformal towers associated to spin $0/\frac{1}{2}$ primary WZW fields as marked in Fig.~\ref{fig:ES}. Moreover, the agreement with the CFT level counting becomes better for larger $N_v$.  
Further details regarding computation of ES and finite-size dependence of its two topological sectors can be found in Appendix~\footnote{In the supplemental materials, we provide more detailed results, including the finite-size analysis of the ES, CSL characteristics from the unrestricted simulations and more signatures of the phase transition. }.

\begin{figure}[t]
\centering
\includegraphics[width=\columnwidth]{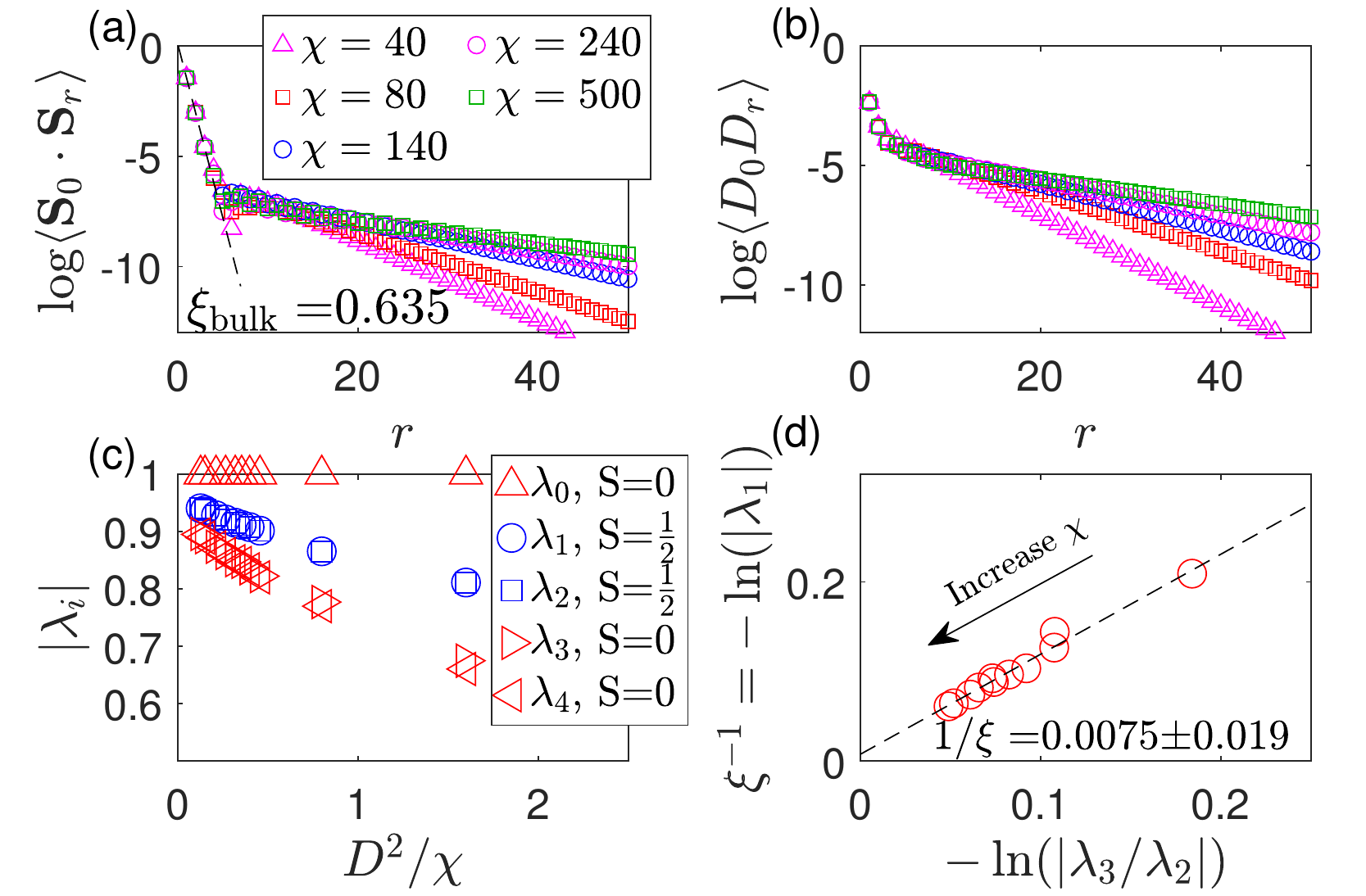}
\caption{Bulk correlations in the optimized $\mathrm{A}_1+i\mathrm{A}_2$ chiral iPESS ($D=8$) at $\theta=0.2\pi$. (a) and (b) show connected spin and dimer correlation functions where $D_i=\mathbf{S}_i\cdot\mathbf{S_{i+1}}$ with increasing CTMRG environment dimension $\chi$. (c) Normalized SU$(2)$ multiplet eigenvalues $\lambda_i$ of the transfer matrix. The gap of the spectrum $|\lambda_0|-|\lambda_1|$ scales to zero in the large $\chi$ limit. (d) Scaling of the leading correlation length $\xi$.}
\label{fig:correlation_fun}
\end{figure}

After analysis of the boundary properties of our CSL iPESS we turn to a second important aspect, the bulk correlations. In all previous studies, done exclusively on the square lattice, CSLs described by iPEPS display weak long-range tails in bulk correlations~\cite{poilblanc2015chiral,systematic2016,poilblanc2017,chen2018non,chen2020,chen2021abelian,hasik2022}, contrary to the expectation for a gapped CSL phase. We show the same behaviour arises also in bulk correlations on the kagome lattice, depicted in Fig.~\ref{fig:correlation_fun} for $D=8$ state.
Spin and dimer correlation functions clearly show two distinct regimes. At short distance $r\le 4$ the spin correlations decay exponentially with very short correlation length $\xi_{\rm bulk}\approx 0.6$, which is believed to be related to the bulk gap of the true ground state~\cite{chen2018non,hasik2022}. At longer distances a long-range tail of tiny magnitude ($\approx10^{-6}$) emerges. Its decay becomes slower than any exponential function as its correlation length increases with $\chi$. To confirm this we analyze the transfer matrix spectrum~\cite{nishino1996} in Fig.~\ref{fig:correlation_fun} (c, d), which demonstrates that by increasing $\chi$, the gap in the spectrum $|\lambda_0|-|\lambda_1|$ (normalization $|\lambda_0|=1$ is used) vanishes and the leading correlation length $\xi=-1/\mathrm{ln}|\lambda_1|$ (associated with the tail) diverges. The dimer correlations behave similarly.
These long-range tails indicate the state supports some form of long-range bulk correlations, which still could decay faster than any power law (see discussion in Ref.~\onlinecite{poilblanc2017}). 
These features come as a direct consequence of the no-go theorem~\cite{wahl2013projected,yang2015chiral,dubail2015tensor} formally preventing the exact finite-$D$ TN representation of gapped chiral topological states. This can be physically understood from the PEPS bulk-edge correspondence: The entanglement Hamiltonian necessary for harboring an ideal gapless chiral ES, strictly speaking, must be of infinite range which, in turn, implies an infinite bulk correlation length~\cite{poilblanc2012topological,cirac2011entanglement}. These features closely resemble those observed in the iPEPS representations of CSL on the square lattice~\cite{poilblanc2017,hasik2022}, strongly suggesting a universal finite-$D$ artifact in all TN ansatze of spin-1/2 CSL.

Good variational energies, sharply defined chiral ES branches, and exponentially decaying correlations at short distances clearly demonstrate the ability of  iPESS developed here to simulate CSL phases on the kagome lattice. We conjecture that the unphysical weak  long-distance gossamer tail will eventually vanish upon increasing $D$. In the Appendix we show the same CSL features are obtained from iPESS simulations without  symmetries, which indicates that the CSL phase in the iPESS representation is not fine-tuned. 

\begin{figure}[t]
\centering
\includegraphics[width=\columnwidth]{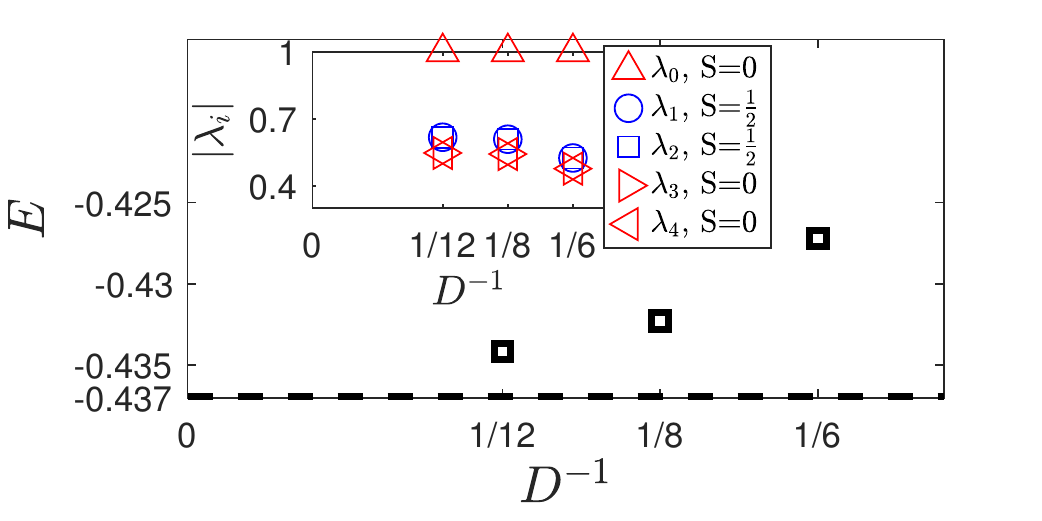}
\caption{The variational energies of the non-chiral $t_{\mathrm{A}_1}^{\mathrm{I}}+t_{\mathrm{A}_2}^{\mathrm{II}}$ ansatz at the KHA point. The dashed line shows the best energy $-0.4369$  in Ref.~\cite{KHA2017gap} with $D=29$ ($D^*=12$). Inset shows the transfer matrix eigenvalues for different $D$.}
\label{fig:KHA}
\end{figure}

\emph{The KHA point and the transition to the CSL phase.---} We now construct the optimal \emph{point group symmetric} ansatz at the KHA $J_\chi=0$ point and investigate the possible occurrence of spontaneous time-reversal symmetry breaking induced by the proximity to the CSL phase~\cite{he2015distinct,he2014chiral,gong2014emergent,sun2022possible}. Starting the optimizations from  random $\mathrm{A}_1+i\mathrm{A}_2$ chiral states 
we find they always flow to the non-chiral $t_{\mathrm{A}_1}^{\mathrm{I}}+t_{\mathrm{A}_2}^{\mathrm{II}}$ ansatz (the same as the $D=3$ RVB state), which is a subset of the complete family of chiral ansatze.
This suggests that spontaneous time reversal symmetry breaking does not occur at the KHA point. 
As shown in Fig.~\ref{fig:KHA}, the variational energies of our finite-$D$ non-chiral ansatze are consistent with the finite $D$ energies 
obtained in Ref.~\cite{KHA2017gap} where only SU$(2)$ symmetry is imposed, hence validating our fully symmetric ansatz. At currently accessible $D=6,8,12$ the resulting iPESS is gapped and has $\mathbb{Z}_2$ topological order~\cite{KHA2017gap} (in contrast to the Gutzwiller-projected Dirac liquid of similar energy~\cite{iqbal2013}). The large gap seen in the transfer matrix spectrum shown in the inset of Fig.~\ref{fig:KHA} implies that, at current $D$, the correlation length is short and a larger $D$ may still be needed to approach the true ground state. However, our new knowledge of the exact form of the optimal ansatz is expected to greatly speed up the variational optimization at larger $D$.

\begin{figure}[t]
\centering
\includegraphics[width=\columnwidth]{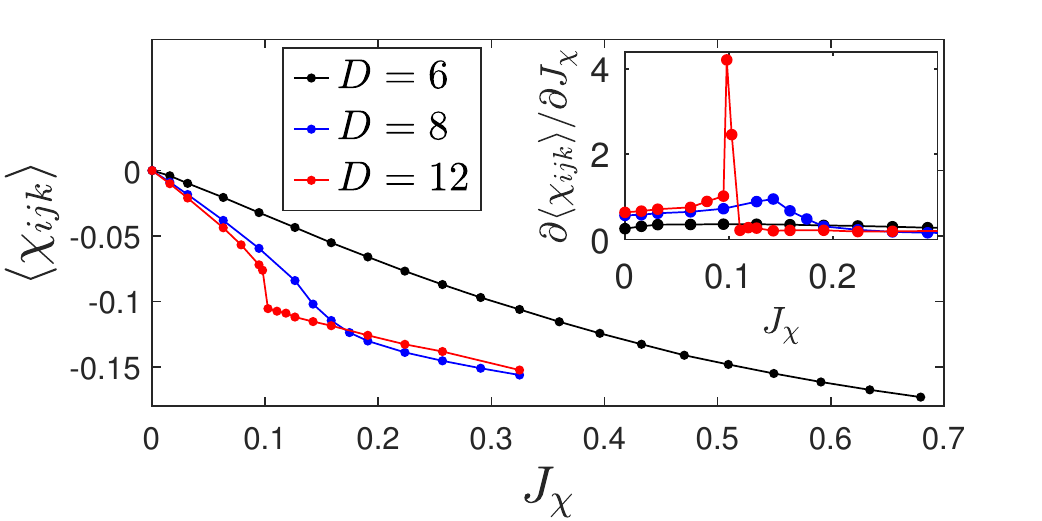}
\caption{The observable $\langle \chi_{ijk} \rangle$ of the optimized chiral states. Here $\chi=60$ is used. Inset shows the first order derivatives.  }
\label{fig:chiral_order}
\end{figure}

To investigate the transition between the CSL phase at large $J_{\chi}$ and the non-chiral spin liquid phase at $J_{\chi}=0$, we look at the scalar chirality  $\langle\chi_{ijk}\rangle$ in the optimized ansatz, as shown in Fig.~\ref{fig:chiral_order}. A linear growth of $|\langle\chi_{ijk}\rangle|$ is observed at small $J_\chi$, consistent with linear response theory and a finite value of the chiral susceptibility $\chi_{\rm chiral}=-\partial\langle\chi_{ijk}\rangle/\partial J_\chi|_{J_\chi=0}$ (measured by the slope at $J_\chi\rightarrow 0$).
For $D\geq8$ a singularity builds up at finite $J_\chi$, as reflected more clearly by the behavior of the first (numerical) derivative of $|\langle\chi_{ijk}\rangle|$ w.r.t. $J_\chi$
suggesting a first order transition around $J_\chi\simeq 0.1$, in agreement with the DMRG result in Ref.~\cite{haghshenas2019single}. The transition detected from the jump of $\langle\chi_{ijk}\rangle$ is also confirmed by behaviors of the variational parameters and the ES, see the Appendix. Note the value of the critical point may be overestimated since the gap of the variational state at the KHA point decreases with increasing $D$ making the system more susceptible to stabilize the CSL phase.     

\emph{Conclusion and outlook.---}In this paper, we demonstrated that the CSL ground state of the chiral Heisenberg model on the kagome lattice can be represented faithfully by the chiral iPESS ansatz. By classifying elementary tensors with optimal virtual spaces using SU$(2)$ and point group symmetries, we constructed an enlarged family of chiral states for numerical simulations (including a non-chiral subset) and studied the transition induced by the scalar chirality term between the non-chiral spin liquid phase and the CSL phase. In contrast to the former, the latter is characterized by a well-defined chiral edge mode consistent with the SU($2$)$_1$ conformal field theory. A long-range tail (of very small weight) in the spin-spin correlation is found, as also encountered in the square lattice CSL analog~\cite{poilblanc2017,hasik2022}, suggesting universality in finite-$D$ artifacts of TN representation of CSL. 
The non-chiral ansatz discovered here paves the way for the symmetric variational optimization and the study of the excitation spectrum at the KHA point. Also, it would be interesting to apply our chiral ansatz to explore spontaneous time reversal symmetry breaking in the model with longer range couplings~\cite{he2015distinct,he2014chiral,gong2014emergent,sun2022possible}. 

\emph{Acknowledgements.---}We thank S. Capponi for providing the ED data with $N=36$ in Fig.~\ref{fig:energy_0d2pi}. We implement non-abelian symmetries using the QSpace tensor library developed by A. Weichselbaum~\cite{Weichselbaum2012, Weichselbaum2020} and the TensorKit.jl package~\cite{tensorkit}. We acknowledge J. von Delft for providing part of the computational resource. This work was granted access to the HPC resources of CALMIP center under the allocation 2017-P1231. J.-Y.C. acknowledges support by the Deutsche Forschungsgemeinschaft (DFG, German Research Foundation) through CRC 183 (project B01) and the Sun Yat-sen University through a startup grant. This work was also supported by the TNTOP ANR-18-CE30-0026-01 grant awarded by the French Research Council and the European Research Council (ERC) under the European Union's Horizon 2020 research and innovation programme (grant agreement No 101001604).

\bibliography{kagome_CSL.bib}

\begin{thebibliography}{53}%
\makeatletter
\providecommand \@ifxundefined [1]{%
 \@ifx{#1\undefined}
}%
\providecommand \@ifnum [1]{%
 \ifnum #1\expandafter \@firstoftwo
 \else \expandafter \@secondoftwo
 \fi
}%
\providecommand \@ifx [1]{%
 \ifx #1\expandafter \@firstoftwo
 \else \expandafter \@secondoftwo
 \fi
}%
\providecommand \natexlab [1]{#1}%
\providecommand \enquote  [1]{``#1''}%
\providecommand \bibnamefont  [1]{#1}%
\providecommand \bibfnamefont [1]{#1}%
\providecommand \citenamefont [1]{#1}%
\providecommand \href@noop [0]{\@secondoftwo}%
\providecommand \href [0]{\begingroup \@sanitize@url \@href}%
\providecommand \@href[1]{\@@startlink{#1}\@@href}%
\providecommand \@@href[1]{\endgroup#1\@@endlink}%
\providecommand \@sanitize@url [0]{\catcode `\\12\catcode `\$12\catcode
  `\&12\catcode `\#12\catcode `\^12\catcode `\_12\catcode `\%12\relax}%
\providecommand \@@startlink[1]{}%
\providecommand \@@endlink[0]{}%
\providecommand \url  [0]{\begingroup\@sanitize@url \@url }%
\providecommand \@url [1]{\endgroup\@href {#1}{\urlprefix }}%
\providecommand \urlprefix  [0]{URL }%
\providecommand \Eprint [0]{\href }%
\providecommand \doibase [0]{https://doi.org/}%
\providecommand \selectlanguage [0]{\@gobble}%
\providecommand \bibinfo  [0]{\@secondoftwo}%
\providecommand \bibfield  [0]{\@secondoftwo}%
\providecommand \translation [1]{[#1]}%
\providecommand \BibitemOpen [0]{}%
\providecommand \bibitemStop [0]{}%
\providecommand \bibitemNoStop [0]{.\EOS\space}%
\providecommand \EOS [0]{\spacefactor3000\relax}%
\providecommand \BibitemShut  [1]{\csname bibitem#1\endcsname}%
\let\auto@bib@innerbib\@empty
\bibitem [{\citenamefont {Savary}\ and\ \citenamefont
  {Balents}(2016)}]{savary2016quantum}%
  \BibitemOpen
  \bibfield  {author} {\bibinfo {author} {\bibfnamefont {L.}~\bibnamefont
  {Savary}}\ and\ \bibinfo {author} {\bibfnamefont {L.}~\bibnamefont
  {Balents}},\ }\bibfield  {title} {\bibinfo {title} {Quantum spin liquids: a
  review},\ }\href@noop {} {\bibfield  {journal} {\bibinfo  {journal} {Reports
  on Progress in Physics}\ }\textbf {\bibinfo {volume} {80}},\ \bibinfo {pages}
  {016502} (\bibinfo {year} {2016})}\BibitemShut {NoStop}%
\bibitem [{\citenamefont {Zhou}\ \emph {et~al.}(2017)\citenamefont {Zhou},
  \citenamefont {Kanoda},\ and\ \citenamefont {Ng}}]{zhou2017quantum}%
  \BibitemOpen
  \bibfield  {author} {\bibinfo {author} {\bibfnamefont {Y.}~\bibnamefont
  {Zhou}}, \bibinfo {author} {\bibfnamefont {K.}~\bibnamefont {Kanoda}},\ and\
  \bibinfo {author} {\bibfnamefont {T.-K.}\ \bibnamefont {Ng}},\ }\bibfield
  {title} {\bibinfo {title} {Quantum spin liquid states},\ }\href@noop {}
  {\bibfield  {journal} {\bibinfo  {journal} {Rev. Mod. Phys.}\ }\textbf
  {\bibinfo {volume} {89}},\ \bibinfo {pages} {025003} (\bibinfo {year}
  {2017})}\BibitemShut {NoStop}%
\bibitem [{\citenamefont {Ran}\ \emph {et~al.}(2007)\citenamefont {Ran},
  \citenamefont {Hermele}, \citenamefont {Lee},\ and\ \citenamefont
  {Wen}}]{ran2007projected}%
  \BibitemOpen
  \bibfield  {author} {\bibinfo {author} {\bibfnamefont {Y.}~\bibnamefont
  {Ran}}, \bibinfo {author} {\bibfnamefont {M.}~\bibnamefont {Hermele}},
  \bibinfo {author} {\bibfnamefont {P.~A.}\ \bibnamefont {Lee}},\ and\ \bibinfo
  {author} {\bibfnamefont {X.-G.}\ \bibnamefont {Wen}},\ }\bibfield  {title}
  {\bibinfo {title} {Projected-wave-function study of the spin-1/2 {H}eisenberg
  model on the kagom{\'e} lattice},\ }\href@noop {} {\bibfield  {journal}
  {\bibinfo  {journal} {Phys. Rev. Lett.}\ }\textbf {\bibinfo {volume} {98}},\
  \bibinfo {pages} {117205} (\bibinfo {year} {2007})}\BibitemShut {NoStop}%
\bibitem [{\citenamefont {Sachdev}(1992)}]{sachdev1992kagome}%
  \BibitemOpen
  \bibfield  {author} {\bibinfo {author} {\bibfnamefont {S.}~\bibnamefont
  {Sachdev}},\ }\bibfield  {title} {\bibinfo {title} {Kagom{\'e}-and
  triangular-lattice {H}eisenberg antiferromagnets: Ordering from quantum
  fluctuations and quantum-disordered ground states with unconfined bosonic
  spinons},\ }\href@noop {} {\bibfield  {journal} {\bibinfo  {journal} {Phys.
  Rev. B}\ }\textbf {\bibinfo {volume} {45}},\ \bibinfo {pages} {12377}
  (\bibinfo {year} {1992})}\BibitemShut {NoStop}%
\bibitem [{\citenamefont {Bauer}\ \emph {et~al.}(2014)\citenamefont {Bauer},
  \citenamefont {Cincio}, \citenamefont {Keller}, \citenamefont {Dolfi},
  \citenamefont {Vidal}, \citenamefont {Trebst},\ and\ \citenamefont
  {Ludwig}}]{kagomeCSL2014}%
  \BibitemOpen
  \bibfield  {author} {\bibinfo {author} {\bibfnamefont {B.}~\bibnamefont
  {Bauer}}, \bibinfo {author} {\bibfnamefont {L.}~\bibnamefont {Cincio}},
  \bibinfo {author} {\bibfnamefont {B.~P.}\ \bibnamefont {Keller}}, \bibinfo
  {author} {\bibfnamefont {M.}~\bibnamefont {Dolfi}}, \bibinfo {author}
  {\bibfnamefont {G.}~\bibnamefont {Vidal}}, \bibinfo {author} {\bibfnamefont
  {S.}~\bibnamefont {Trebst}},\ and\ \bibinfo {author} {\bibfnamefont {A.~W.}\
  \bibnamefont {Ludwig}},\ }\bibfield  {title} {\bibinfo {title} {Chiral spin
  liquid and emergent anyons in a kagome lattice mott insulator},\ }\href@noop
  {} {\bibfield  {journal} {\bibinfo  {journal} {Nat. Commun.}\ }\textbf
  {\bibinfo {volume} {5}},\ \bibinfo {pages} {1} (\bibinfo {year}
  {2014})}\BibitemShut {NoStop}%
\bibitem [{\citenamefont {Gong}\ \emph {et~al.}(2014)\citenamefont {Gong},
  \citenamefont {Zhu},\ and\ \citenamefont {Sheng}}]{gong2014emergent}%
  \BibitemOpen
  \bibfield  {author} {\bibinfo {author} {\bibfnamefont {S.-S.}\ \bibnamefont
  {Gong}}, \bibinfo {author} {\bibfnamefont {W.}~\bibnamefont {Zhu}},\ and\
  \bibinfo {author} {\bibfnamefont {D.}~\bibnamefont {Sheng}},\ }\bibfield
  {title} {\bibinfo {title} {Emergent chiral spin liquid: Fractional quantum
  hall effect in a kagome {H}eisenberg model},\ }\href@noop {} {\bibfield
  {journal} {\bibinfo  {journal} {Sci. Rep.}\ }\textbf {\bibinfo {volume}
  {4}},\ \bibinfo {pages} {1} (\bibinfo {year} {2014})}\BibitemShut {NoStop}%
\bibitem [{\citenamefont {Tsui}\ \emph {et~al.}(1982)\citenamefont {Tsui},
  \citenamefont {Stormer},\ and\ \citenamefont {Gossard}}]{tsui1982two}%
  \BibitemOpen
  \bibfield  {author} {\bibinfo {author} {\bibfnamefont {D.~C.}\ \bibnamefont
  {Tsui}}, \bibinfo {author} {\bibfnamefont {H.~L.}\ \bibnamefont {Stormer}},\
  and\ \bibinfo {author} {\bibfnamefont {A.~C.}\ \bibnamefont {Gossard}},\
  }\bibfield  {title} {\bibinfo {title} {Two-dimensional magnetotransport in
  the extreme quantum limit},\ }\href@noop {} {\bibfield  {journal} {\bibinfo
  {journal} {Phys. Rev. Lett.}\ }\textbf {\bibinfo {volume} {48}},\ \bibinfo
  {pages} {1559} (\bibinfo {year} {1982})}\BibitemShut {NoStop}%
\bibitem [{\citenamefont {Halperin}(1984)}]{halperin1984statistics}%
  \BibitemOpen
  \bibfield  {author} {\bibinfo {author} {\bibfnamefont {B.~I.}\ \bibnamefont
  {Halperin}},\ }\bibfield  {title} {\bibinfo {title} {Statistics of
  quasiparticles and the hierarchy of fractional quantized hall states},\
  }\href@noop {} {\bibfield  {journal} {\bibinfo  {journal} {Phys. Rev. Lett.}\
  }\textbf {\bibinfo {volume} {52}},\ \bibinfo {pages} {1583} (\bibinfo {year}
  {1984})}\BibitemShut {NoStop}%
\bibitem [{\citenamefont {Wen}(1991)}]{wen1991gapless}%
  \BibitemOpen
  \bibfield  {author} {\bibinfo {author} {\bibfnamefont {X.-G.}\ \bibnamefont
  {Wen}},\ }\bibfield  {title} {\bibinfo {title} {Gapless boundary excitations
  in the quantum hall states and in the chiral spin states},\ }\href@noop {}
  {\bibfield  {journal} {\bibinfo  {journal} {Phys. Rev. B}\ }\textbf {\bibinfo
  {volume} {43}},\ \bibinfo {pages} {11025} (\bibinfo {year}
  {1991})}\BibitemShut {NoStop}%
\bibitem [{\citenamefont {Verstraete}\ \emph {et~al.}(2008)\citenamefont
  {Verstraete}, \citenamefont {Murg},\ and\ \citenamefont
  {Cirac}}]{verstraete2008matrix}%
  \BibitemOpen
  \bibfield  {author} {\bibinfo {author} {\bibfnamefont {F.}~\bibnamefont
  {Verstraete}}, \bibinfo {author} {\bibfnamefont {V.}~\bibnamefont {Murg}},\
  and\ \bibinfo {author} {\bibfnamefont {J.~I.}\ \bibnamefont {Cirac}},\
  }\bibfield  {title} {\bibinfo {title} {Matrix product states, projected
  entangled pair states, and variational renormalization group methods for
  quantum spin systems},\ }\href@noop {} {\bibfield  {journal} {\bibinfo
  {journal} {Adv. Phys.}\ }\textbf {\bibinfo {volume} {57}},\ \bibinfo {pages}
  {143} (\bibinfo {year} {2008})}\BibitemShut {NoStop}%
\bibitem [{\citenamefont {Verstraete}\ \emph {et~al.}(2006)\citenamefont
  {Verstraete}, \citenamefont {Wolf}, \citenamefont {Perez-Garcia},\ and\
  \citenamefont {Cirac}}]{toriccode2006}%
  \BibitemOpen
  \bibfield  {author} {\bibinfo {author} {\bibfnamefont {F.}~\bibnamefont
  {Verstraete}}, \bibinfo {author} {\bibfnamefont {M.~M.}\ \bibnamefont
  {Wolf}}, \bibinfo {author} {\bibfnamefont {D.}~\bibnamefont {Perez-Garcia}},\
  and\ \bibinfo {author} {\bibfnamefont {J.~I.}\ \bibnamefont {Cirac}},\
  }\bibfield  {title} {\bibinfo {title} {Criticality, the area law, and the
  computational power of projected entangled pair states},\ }\href@noop {}
  {\bibfield  {journal} {\bibinfo  {journal} {Phys. Rev. Lett.}\ }\textbf
  {\bibinfo {volume} {96}},\ \bibinfo {pages} {220601} (\bibinfo {year}
  {2006})}\BibitemShut {NoStop}%
\bibitem [{\citenamefont {Schuch}\ \emph {et~al.}(2010)\citenamefont {Schuch},
  \citenamefont {Cirac},\ and\ \citenamefont
  {P{\'e}rez-Garc{\'\i}a}}]{toriccode2009}%
  \BibitemOpen
  \bibfield  {author} {\bibinfo {author} {\bibfnamefont {N.}~\bibnamefont
  {Schuch}}, \bibinfo {author} {\bibfnamefont {I.}~\bibnamefont {Cirac}},\ and\
  \bibinfo {author} {\bibfnamefont {D.}~\bibnamefont {P{\'e}rez-Garc{\'\i}a}},\
  }\bibfield  {title} {\bibinfo {title} {{PEPS} as ground states: Degeneracy
  and topology},\ }\href@noop {} {\bibfield  {journal} {\bibinfo  {journal}
  {Ann. Phys.}\ }\textbf {\bibinfo {volume} {325}},\ \bibinfo {pages} {2153}
  (\bibinfo {year} {2010})}\BibitemShut {NoStop}%
\bibitem [{\citenamefont {Gu}\ \emph {et~al.}(2009)\citenamefont {Gu},
  \citenamefont {Levin}, \citenamefont {Swingle},\ and\ \citenamefont
  {Wen}}]{stringnetnet2009}%
  \BibitemOpen
  \bibfield  {author} {\bibinfo {author} {\bibfnamefont {Z.-C.}\ \bibnamefont
  {Gu}}, \bibinfo {author} {\bibfnamefont {M.}~\bibnamefont {Levin}}, \bibinfo
  {author} {\bibfnamefont {B.}~\bibnamefont {Swingle}},\ and\ \bibinfo {author}
  {\bibfnamefont {X.-G.}\ \bibnamefont {Wen}},\ }\bibfield  {title} {\bibinfo
  {title} {Tensor-product representations for string-net condensed states},\
  }\href@noop {} {\bibfield  {journal} {\bibinfo  {journal} {Phys. Rev. B}\
  }\textbf {\bibinfo {volume} {79}},\ \bibinfo {pages} {085118} (\bibinfo
  {year} {2009})}\BibitemShut {NoStop}%
\bibitem [{\citenamefont {Poilblanc}\ \emph {et~al.}(2012)\citenamefont
  {Poilblanc}, \citenamefont {Schuch}, \citenamefont {P{\'e}rez-Garc{\'\i}a},\
  and\ \citenamefont {Cirac}}]{poilblanc2012topological}%
  \BibitemOpen
  \bibfield  {author} {\bibinfo {author} {\bibfnamefont {D.}~\bibnamefont
  {Poilblanc}}, \bibinfo {author} {\bibfnamefont {N.}~\bibnamefont {Schuch}},
  \bibinfo {author} {\bibfnamefont {D.}~\bibnamefont {P{\'e}rez-Garc{\'\i}a}},\
  and\ \bibinfo {author} {\bibfnamefont {J.~I.}\ \bibnamefont {Cirac}},\
  }\bibfield  {title} {\bibinfo {title} {Topological and entanglement
  properties of resonating valence bond wave functions},\ }\href@noop {}
  {\bibfield  {journal} {\bibinfo  {journal} {Phys. Rev. B}\ }\textbf {\bibinfo
  {volume} {86}},\ \bibinfo {pages} {014404} (\bibinfo {year}
  {2012})}\BibitemShut {NoStop}%
\bibitem [{\citenamefont {Iqbal}\ \emph {et~al.}(2013)\citenamefont {Iqbal},
  \citenamefont {Becca}, \citenamefont {Sorella},\ and\ \citenamefont
  {Poilblanc}}]{iqbal2013}%
  \BibitemOpen
  \bibfield  {author} {\bibinfo {author} {\bibfnamefont {Y.}~\bibnamefont
  {Iqbal}}, \bibinfo {author} {\bibfnamefont {F.}~\bibnamefont {Becca}},
  \bibinfo {author} {\bibfnamefont {S.}~\bibnamefont {Sorella}},\ and\ \bibinfo
  {author} {\bibfnamefont {D.}~\bibnamefont {Poilblanc}},\ }\bibfield  {title}
  {\bibinfo {title} {Gapless spin-liquid phase in the kagome spin-$\frac{1}{2}$
  {H}eisenberg antiferromagnet},\ }\href
  {https://doi.org/10.1103/PhysRevB.87.060405} {\bibfield  {journal} {\bibinfo
  {journal} {Phys. Rev. B}\ }\textbf {\bibinfo {volume} {87}},\ \bibinfo
  {pages} {060405} (\bibinfo {year} {2013})}\BibitemShut {NoStop}%
\bibitem [{\citenamefont {Mei}\ \emph {et~al.}(2017)\citenamefont {Mei},
  \citenamefont {Chen}, \citenamefont {He},\ and\ \citenamefont
  {Wen}}]{KHA2017gap}%
  \BibitemOpen
  \bibfield  {author} {\bibinfo {author} {\bibfnamefont {J.-W.}\ \bibnamefont
  {Mei}}, \bibinfo {author} {\bibfnamefont {J.-Y.}\ \bibnamefont {Chen}},
  \bibinfo {author} {\bibfnamefont {H.}~\bibnamefont {He}},\ and\ \bibinfo
  {author} {\bibfnamefont {X.-G.}\ \bibnamefont {Wen}},\ }\bibfield  {title}
  {\bibinfo {title} {Gapped spin liquid with ${Z}_2$ topological order for the
  kagome {H}eisenberg model},\ }\href@noop {} {\bibfield  {journal} {\bibinfo
  {journal} {Phys. Rev. B}\ }\textbf {\bibinfo {volume} {95}},\ \bibinfo
  {pages} {235107} (\bibinfo {year} {2017})}\BibitemShut {NoStop}%
\bibitem [{\citenamefont {Liao}\ \emph {et~al.}(2017)\citenamefont {Liao},
  \citenamefont {Xie}, \citenamefont {Chen}, \citenamefont {Liu}, \citenamefont
  {Xie}, \citenamefont {Huang}, \citenamefont {Normand},\ and\ \citenamefont
  {Xiang}}]{KHA2017gapless}%
  \BibitemOpen
  \bibfield  {author} {\bibinfo {author} {\bibfnamefont {H.-J.}\ \bibnamefont
  {Liao}}, \bibinfo {author} {\bibfnamefont {Z.-Y.}\ \bibnamefont {Xie}},
  \bibinfo {author} {\bibfnamefont {J.}~\bibnamefont {Chen}}, \bibinfo {author}
  {\bibfnamefont {Z.-Y.}\ \bibnamefont {Liu}}, \bibinfo {author} {\bibfnamefont
  {H.-D.}\ \bibnamefont {Xie}}, \bibinfo {author} {\bibfnamefont {R.-Z.}\
  \bibnamefont {Huang}}, \bibinfo {author} {\bibfnamefont {B.}~\bibnamefont
  {Normand}},\ and\ \bibinfo {author} {\bibfnamefont {T.}~\bibnamefont
  {Xiang}},\ }\bibfield  {title} {\bibinfo {title} {Gapless spin-liquid ground
  state in the s= 1/2 kagome antiferromagnet},\ }\href@noop {} {\bibfield
  {journal} {\bibinfo  {journal} {Phys. Rev. Lett.}\ }\textbf {\bibinfo
  {volume} {118}},\ \bibinfo {pages} {137202} (\bibinfo {year}
  {2017})}\BibitemShut {NoStop}%
\bibitem [{\citenamefont {Schuch}\ \emph {et~al.}(2013)\citenamefont {Schuch},
  \citenamefont {Poilblanc}, \citenamefont {Cirac},\ and\ \citenamefont
  {Perez-Garcia}}]{formalism2013}%
  \BibitemOpen
  \bibfield  {author} {\bibinfo {author} {\bibfnamefont {N.}~\bibnamefont
  {Schuch}}, \bibinfo {author} {\bibfnamefont {D.}~\bibnamefont {Poilblanc}},
  \bibinfo {author} {\bibfnamefont {J.~I.}\ \bibnamefont {Cirac}},\ and\
  \bibinfo {author} {\bibfnamefont {D.}~\bibnamefont {Perez-Garcia}},\
  }\bibfield  {title} {\bibinfo {title} {Topological order in the projected
  entangled-pair states formalism: Transfer operator and boundary
  hamiltonians},\ }\href@noop {} {\bibfield  {journal} {\bibinfo  {journal}
  {Phys. Rev. Lett.}\ }\textbf {\bibinfo {volume} {111}},\ \bibinfo {pages}
  {090501} (\bibinfo {year} {2013})}\BibitemShut {NoStop}%
\bibitem [{\citenamefont {Wahl}\ \emph {et~al.}(2013)\citenamefont {Wahl},
  \citenamefont {Tu}, \citenamefont {Schuch},\ and\ \citenamefont
  {Cirac}}]{wahl2013projected}%
  \BibitemOpen
  \bibfield  {author} {\bibinfo {author} {\bibfnamefont {T.~B.}\ \bibnamefont
  {Wahl}}, \bibinfo {author} {\bibfnamefont {H.-H.}\ \bibnamefont {Tu}},
  \bibinfo {author} {\bibfnamefont {N.}~\bibnamefont {Schuch}},\ and\ \bibinfo
  {author} {\bibfnamefont {J.~I.}\ \bibnamefont {Cirac}},\ }\bibfield  {title}
  {\bibinfo {title} {Projected entangled-pair states can describe chiral
  topological states},\ }\href@noop {} {\bibfield  {journal} {\bibinfo
  {journal} {Phys. Rev. Lett.}\ }\textbf {\bibinfo {volume} {111}},\ \bibinfo
  {pages} {236805} (\bibinfo {year} {2013})}\BibitemShut {NoStop}%
\bibitem [{\citenamefont {Yang}\ \emph {et~al.}(2015)\citenamefont {Yang},
  \citenamefont {Wahl}, \citenamefont {Tu}, \citenamefont {Schuch},\ and\
  \citenamefont {Cirac}}]{yang2015chiral}%
  \BibitemOpen
  \bibfield  {author} {\bibinfo {author} {\bibfnamefont {S.}~\bibnamefont
  {Yang}}, \bibinfo {author} {\bibfnamefont {T.~B.}\ \bibnamefont {Wahl}},
  \bibinfo {author} {\bibfnamefont {H.-H.}\ \bibnamefont {Tu}}, \bibinfo
  {author} {\bibfnamefont {N.}~\bibnamefont {Schuch}},\ and\ \bibinfo {author}
  {\bibfnamefont {J.~I.}\ \bibnamefont {Cirac}},\ }\bibfield  {title} {\bibinfo
  {title} {Chiral projected entangled-pair state with topological order},\
  }\href@noop {} {\bibfield  {journal} {\bibinfo  {journal} {Phys. Rev. Lett.}\
  }\textbf {\bibinfo {volume} {114}},\ \bibinfo {pages} {106803} (\bibinfo
  {year} {2015})}\BibitemShut {NoStop}%
\bibitem [{\citenamefont {Dubail}\ and\ \citenamefont
  {Read}(2015)}]{dubail2015tensor}%
  \BibitemOpen
  \bibfield  {author} {\bibinfo {author} {\bibfnamefont {J.}~\bibnamefont
  {Dubail}}\ and\ \bibinfo {author} {\bibfnamefont {N.}~\bibnamefont {Read}},\
  }\bibfield  {title} {\bibinfo {title} {Tensor network trial states for chiral
  topological phases in two dimensions and a no-go theorem in any dimension},\
  }\href@noop {} {\bibfield  {journal} {\bibinfo  {journal} {Phys. Rev. B}\
  }\textbf {\bibinfo {volume} {92}},\ \bibinfo {pages} {205307} (\bibinfo
  {year} {2015})}\BibitemShut {NoStop}%
\bibitem [{\citenamefont {Poilblanc}\ \emph {et~al.}(2015)\citenamefont
  {Poilblanc}, \citenamefont {Cirac},\ and\ \citenamefont
  {Schuch}}]{poilblanc2015chiral}%
  \BibitemOpen
  \bibfield  {author} {\bibinfo {author} {\bibfnamefont {D.}~\bibnamefont
  {Poilblanc}}, \bibinfo {author} {\bibfnamefont {J.~I.}\ \bibnamefont
  {Cirac}},\ and\ \bibinfo {author} {\bibfnamefont {N.}~\bibnamefont
  {Schuch}},\ }\bibfield  {title} {\bibinfo {title} {Chiral topological spin
  liquids with projected entangled pair states},\ }\href@noop {} {\bibfield
  {journal} {\bibinfo  {journal} {Phys. Rev. B}\ }\textbf {\bibinfo {volume}
  {91}},\ \bibinfo {pages} {224431} (\bibinfo {year} {2015})}\BibitemShut
  {NoStop}%
\bibitem [{\citenamefont {Mambrini}\ \emph {et~al.}(2016)\citenamefont
  {Mambrini}, \citenamefont {Or{\'u}s},\ and\ \citenamefont
  {Poilblanc}}]{systematic2016}%
  \BibitemOpen
  \bibfield  {author} {\bibinfo {author} {\bibfnamefont {M.}~\bibnamefont
  {Mambrini}}, \bibinfo {author} {\bibfnamefont {R.}~\bibnamefont {Or{\'u}s}},\
  and\ \bibinfo {author} {\bibfnamefont {D.}~\bibnamefont {Poilblanc}},\
  }\bibfield  {title} {\bibinfo {title} {Systematic construction of spin
  liquids on the square lattice from tensor networks with {SU}$(2)$ symmetry},\
  }\href@noop {} {\bibfield  {journal} {\bibinfo  {journal} {Phys. Rev. B}\
  }\textbf {\bibinfo {volume} {94}},\ \bibinfo {pages} {205124} (\bibinfo
  {year} {2016})}\BibitemShut {NoStop}%
\bibitem [{\citenamefont {Poilblanc}(2017)}]{poilblanc2017}%
  \BibitemOpen
  \bibfield  {author} {\bibinfo {author} {\bibfnamefont {D.}~\bibnamefont
  {Poilblanc}},\ }\bibfield  {title} {\bibinfo {title} {Investigation of the
  chiral antiferromagnetic {H}eisenberg model using projected entangled pair
  states},\ }\href {https://doi.org/10.1103/PhysRevB.96.121118} {\bibfield
  {journal} {\bibinfo  {journal} {Phys. Rev. B}\ }\textbf {\bibinfo {volume}
  {96}},\ \bibinfo {pages} {121118} (\bibinfo {year} {2017})}\BibitemShut
  {NoStop}%
\bibitem [{\citenamefont {Chen}\ \emph {et~al.}(2020)\citenamefont {Chen},
  \citenamefont {Capponi}, \citenamefont {Wietek}, \citenamefont {Mambrini},
  \citenamefont {Schuch},\ and\ \citenamefont {Poilblanc}}]{chen2020}%
  \BibitemOpen
  \bibfield  {author} {\bibinfo {author} {\bibfnamefont {J.-Y.}\ \bibnamefont
  {Chen}}, \bibinfo {author} {\bibfnamefont {S.}~\bibnamefont {Capponi}},
  \bibinfo {author} {\bibfnamefont {A.}~\bibnamefont {Wietek}}, \bibinfo
  {author} {\bibfnamefont {M.}~\bibnamefont {Mambrini}}, \bibinfo {author}
  {\bibfnamefont {N.}~\bibnamefont {Schuch}},\ and\ \bibinfo {author}
  {\bibfnamefont {D.}~\bibnamefont {Poilblanc}},\ }\bibfield  {title} {\bibinfo
  {title} {{SU}(3)$_1$ chiral spin liquid on the square lattice: A view from
  symmetric projected entangled pair states},\ }\href@noop {} {\bibfield
  {journal} {\bibinfo  {journal} {Phys. Rev. Lett.}\ }\textbf {\bibinfo
  {volume} {125}},\ \bibinfo {pages} {017201} (\bibinfo {year}
  {2020})}\BibitemShut {NoStop}%
\bibitem [{\citenamefont {Chen}\ \emph {et~al.}(2021)\citenamefont {Chen},
  \citenamefont {Li}, \citenamefont {Nataf}, \citenamefont {Capponi},
  \citenamefont {Mambrini}, \citenamefont {Totsuka}, \citenamefont {Tu},
  \citenamefont {Weichselbaum}, \citenamefont {von Delft},\ and\ \citenamefont
  {Poilblanc}}]{chen2021abelian}%
  \BibitemOpen
  \bibfield  {author} {\bibinfo {author} {\bibfnamefont {J.-Y.}\ \bibnamefont
  {Chen}}, \bibinfo {author} {\bibfnamefont {J.-W.}\ \bibnamefont {Li}},
  \bibinfo {author} {\bibfnamefont {P.}~\bibnamefont {Nataf}}, \bibinfo
  {author} {\bibfnamefont {S.}~\bibnamefont {Capponi}}, \bibinfo {author}
  {\bibfnamefont {M.}~\bibnamefont {Mambrini}}, \bibinfo {author}
  {\bibfnamefont {K.}~\bibnamefont {Totsuka}}, \bibinfo {author} {\bibfnamefont
  {H.-H.}\ \bibnamefont {Tu}}, \bibinfo {author} {\bibfnamefont
  {A.}~\bibnamefont {Weichselbaum}}, \bibinfo {author} {\bibfnamefont
  {J.}~\bibnamefont {von Delft}},\ and\ \bibinfo {author} {\bibfnamefont
  {D.}~\bibnamefont {Poilblanc}},\ }\bibfield  {title} {\bibinfo {title}
  {Abelian {SU}({N})$_1$ chiral spin liquids on the square lattice},\
  }\href@noop {} {\bibfield  {journal} {\bibinfo  {journal} {Phys. Rev. B}\
  }\textbf {\bibinfo {volume} {104}},\ \bibinfo {pages} {235104} (\bibinfo
  {year} {2021})}\BibitemShut {NoStop}%
\bibitem [{\citenamefont {{Hasik}}\ \emph {et~al.}(2022)\citenamefont
  {{Hasik}}, \citenamefont {{Van Damme}}, \citenamefont {{Poilblanc}},\ and\
  \citenamefont {{Vanderstraeten}}}]{hasik2022}%
  \BibitemOpen
  \bibfield  {author} {\bibinfo {author} {\bibfnamefont {J.}~\bibnamefont
  {{Hasik}}}, \bibinfo {author} {\bibfnamefont {M.}~\bibnamefont {{Van
  Damme}}}, \bibinfo {author} {\bibfnamefont {D.}~\bibnamefont {{Poilblanc}}},\
  and\ \bibinfo {author} {\bibfnamefont {L.}~\bibnamefont {{Vanderstraeten}}},\
  }\bibfield  {title} {\bibinfo {title} {{Simulating Chiral Spin Liquids with
  Projected Entangled-Pair States}},\ }\href@noop {} {\bibfield  {journal}
  {\bibinfo  {journal} {Phys. Rev. Lett.}\ }\textbf {\bibinfo {volume} {129}},\
  \bibinfo {pages} {177201} (\bibinfo {year} {2022})}\BibitemShut {NoStop}%
\bibitem [{\citenamefont {Chen}\ \emph {et~al.}(2018)\citenamefont {Chen},
  \citenamefont {Vanderstraeten}, \citenamefont {Capponi},\ and\ \citenamefont
  {Poilblanc}}]{chen2018non}%
  \BibitemOpen
  \bibfield  {author} {\bibinfo {author} {\bibfnamefont {J.-Y.}\ \bibnamefont
  {Chen}}, \bibinfo {author} {\bibfnamefont {L.}~\bibnamefont
  {Vanderstraeten}}, \bibinfo {author} {\bibfnamefont {S.}~\bibnamefont
  {Capponi}},\ and\ \bibinfo {author} {\bibfnamefont {D.}~\bibnamefont
  {Poilblanc}},\ }\bibfield  {title} {\bibinfo {title} {Non-abelian chiral spin
  liquid in a quantum antiferromagnet revealed by an i{PEPS} study},\
  }\href@noop {} {\bibfield  {journal} {\bibinfo  {journal} {Phys. Rev. B}\
  }\textbf {\bibinfo {volume} {98}},\ \bibinfo {pages} {184409} (\bibinfo
  {year} {2018})}\BibitemShut {NoStop}%
\bibitem [{\citenamefont {Jiang}\ and\ \citenamefont {Ran}(2015)}]{Jiang2015}%
  \BibitemOpen
  \bibfield  {author} {\bibinfo {author} {\bibfnamefont {S.}~\bibnamefont
  {Jiang}}\ and\ \bibinfo {author} {\bibfnamefont {Y.}~\bibnamefont {Ran}},\
  }\bibfield  {title} {\bibinfo {title} {Symmetric tensor networks and
  practical simulation algorithms to sharply identify classes of quantum phases
  distinguishable by short-range physics},\ }\href
  {https://doi.org/10.1103/PhysRevB.92.104414} {\bibfield  {journal} {\bibinfo
  {journal} {Phys. Rev. B}\ }\textbf {\bibinfo {volume} {92}},\ \bibinfo
  {pages} {104414} (\bibinfo {year} {2015})}\BibitemShut {NoStop}%
\bibitem [{\citenamefont {Verstraete}\ and\ \citenamefont
  {Cirac}(2004)}]{verstraete2004renormalization}%
  \BibitemOpen
  \bibfield  {author} {\bibinfo {author} {\bibfnamefont {F.}~\bibnamefont
  {Verstraete}}\ and\ \bibinfo {author} {\bibfnamefont {J.~I.}\ \bibnamefont
  {Cirac}},\ }\bibfield  {title} {\bibinfo {title} {Renormalization algorithms
  for quantum-many body systems in two and higher dimensions},\ }\href@noop {}
  {\bibfield  {journal} {\bibinfo  {journal} {arXiv preprint cond-mat/0407066}\
  } (\bibinfo {year} {2004})}\BibitemShut {NoStop}%
\bibitem [{\citenamefont {Jordan}\ \emph {et~al.}(2008)\citenamefont {Jordan},
  \citenamefont {Or{\'u}s}, \citenamefont {Vidal}, \citenamefont {Verstraete},\
  and\ \citenamefont {Cirac}}]{jordan2008classical}%
  \BibitemOpen
  \bibfield  {author} {\bibinfo {author} {\bibfnamefont {J.}~\bibnamefont
  {Jordan}}, \bibinfo {author} {\bibfnamefont {R.}~\bibnamefont {Or{\'u}s}},
  \bibinfo {author} {\bibfnamefont {G.}~\bibnamefont {Vidal}}, \bibinfo
  {author} {\bibfnamefont {F.}~\bibnamefont {Verstraete}},\ and\ \bibinfo
  {author} {\bibfnamefont {J.~I.}\ \bibnamefont {Cirac}},\ }\bibfield  {title}
  {\bibinfo {title} {Classical simulation of infinite-size quantum lattice
  systems in two spatial dimensions},\ }\href@noop {} {\bibfield  {journal}
  {\bibinfo  {journal} {Phys. Rev. Lett.}\ }\textbf {\bibinfo {volume} {101}},\
  \bibinfo {pages} {250602} (\bibinfo {year} {2008})}\BibitemShut {NoStop}%
\bibitem [{\citenamefont {Schuch}\ \emph {et~al.}(2012)\citenamefont {Schuch},
  \citenamefont {Poilblanc}, \citenamefont {Cirac},\ and\ \citenamefont
  {P\'erez-Garc\'{\i}a}}]{schuch2012}%
  \BibitemOpen
  \bibfield  {author} {\bibinfo {author} {\bibfnamefont {N.}~\bibnamefont
  {Schuch}}, \bibinfo {author} {\bibfnamefont {D.}~\bibnamefont {Poilblanc}},
  \bibinfo {author} {\bibfnamefont {J.~I.}\ \bibnamefont {Cirac}},\ and\
  \bibinfo {author} {\bibfnamefont {D.}~\bibnamefont {P\'erez-Garc\'{\i}a}},\
  }\bibfield  {title} {\bibinfo {title} {Resonating valence bond states in the
  {{PEPS}} formalism},\ }\href {https://doi.org/10.1103/PhysRevB.86.115108}
  {\bibfield  {journal} {\bibinfo  {journal} {Phys. Rev. B}\ }\textbf {\bibinfo
  {volume} {86}},\ \bibinfo {pages} {115108} (\bibinfo {year}
  {2012})}\BibitemShut {NoStop}%
\bibitem [{\citenamefont {Xie}\ \emph {et~al.}(2014)\citenamefont {Xie},
  \citenamefont {Chen}, \citenamefont {Yu}, \citenamefont {Kong}, \citenamefont
  {Normand},\ and\ \citenamefont {Xiang}}]{xie2014tensor}%
  \BibitemOpen
  \bibfield  {author} {\bibinfo {author} {\bibfnamefont {Z.-Y.}\ \bibnamefont
  {Xie}}, \bibinfo {author} {\bibfnamefont {J.}~\bibnamefont {Chen}}, \bibinfo
  {author} {\bibfnamefont {J.}~\bibnamefont {Yu}}, \bibinfo {author}
  {\bibfnamefont {X.}~\bibnamefont {Kong}}, \bibinfo {author} {\bibfnamefont
  {B.}~\bibnamefont {Normand}},\ and\ \bibinfo {author} {\bibfnamefont
  {T.}~\bibnamefont {Xiang}},\ }\bibfield  {title} {\bibinfo {title} {Tensor
  renormalization of quantum many-body systems using projected entangled
  simplex states},\ }\href@noop {} {\bibfield  {journal} {\bibinfo  {journal}
  {Phys. Rev. X}\ }\textbf {\bibinfo {volume} {4}},\ \bibinfo {pages} {011025}
  (\bibinfo {year} {2014})}\BibitemShut {NoStop}%
\bibitem [{\citenamefont {Nishino}\ and\ \citenamefont
  {Okunishi}(1996)}]{ctmrg1}%
  \BibitemOpen
  \bibfield  {author} {\bibinfo {author} {\bibfnamefont {T.}~\bibnamefont
  {Nishino}}\ and\ \bibinfo {author} {\bibfnamefont {K.}~\bibnamefont
  {Okunishi}},\ }\bibfield  {title} {\bibinfo {title} {Corner transfer matrix
  renormalization group method},\ }\href@noop {} {\bibfield  {journal}
  {\bibinfo  {journal} {Journal of the Physical Society of Japan}\ }\textbf
  {\bibinfo {volume} {65}},\ \bibinfo {pages} {891} (\bibinfo {year}
  {1996})}\BibitemShut {NoStop}%
\bibitem [{\citenamefont {Corboz}\ \emph {et~al.}(2014)\citenamefont {Corboz},
  \citenamefont {Rice},\ and\ \citenamefont {Troyer}}]{ctmrg2}%
  \BibitemOpen
  \bibfield  {author} {\bibinfo {author} {\bibfnamefont {P.}~\bibnamefont
  {Corboz}}, \bibinfo {author} {\bibfnamefont {T.~M.}\ \bibnamefont {Rice}},\
  and\ \bibinfo {author} {\bibfnamefont {M.}~\bibnamefont {Troyer}},\
  }\bibfield  {title} {\bibinfo {title} {Competing states in the t-{J} model:
  Uniform d-wave state versus stripe state},\ }\href@noop {} {\bibfield
  {journal} {\bibinfo  {journal} {Phys. Rev. Lett.}\ }\textbf {\bibinfo
  {volume} {113}},\ \bibinfo {pages} {046402} (\bibinfo {year}
  {2014})}\BibitemShut {NoStop}%
\bibitem [{\citenamefont {Jiang}\ \emph {et~al.}(2008)\citenamefont {Jiang},
  \citenamefont {Weng},\ and\ \citenamefont {Xiang}}]{Jiang2008}%
  \BibitemOpen
  \bibfield  {author} {\bibinfo {author} {\bibfnamefont {H.~C.}\ \bibnamefont
  {Jiang}}, \bibinfo {author} {\bibfnamefont {Z.~Y.}\ \bibnamefont {Weng}},\
  and\ \bibinfo {author} {\bibfnamefont {T.}~\bibnamefont {Xiang}},\ }\bibfield
   {title} {\bibinfo {title} {Accurate determination of tensor network state of
  quantum lattice models in two dimensions},\ }\href
  {https://doi.org/10.1103/PhysRevLett.101.090603} {\bibfield  {journal}
  {\bibinfo  {journal} {Phys. Rev. Lett.}\ }\textbf {\bibinfo {volume} {101}},\
  \bibinfo {pages} {090603} (\bibinfo {year} {2008})}\BibitemShut {NoStop}%
\bibitem [{\citenamefont {Jiang}\ \emph {et~al.}(2019)\citenamefont {Jiang},
  \citenamefont {Kim}, \citenamefont {Han},\ and\ \citenamefont
  {Ran}}]{Jiang2019}%
  \BibitemOpen
  \bibfield  {author} {\bibinfo {author} {\bibfnamefont {S.}~\bibnamefont
  {Jiang}}, \bibinfo {author} {\bibfnamefont {P.}~\bibnamefont {Kim}}, \bibinfo
  {author} {\bibfnamefont {J.~H.}\ \bibnamefont {Han}},\ and\ \bibinfo {author}
  {\bibfnamefont {Y.}~\bibnamefont {Ran}},\ }\bibfield  {title} {\bibinfo
  {title} {{Competing Spin Liquid Phases in the S=$\frac{1}{2}$ {H}eisenberg
  Model on the Kagome Lattice}},\ }\href
  {https://doi.org/10.21468/SciPostPhys.7.1.006} {\bibfield  {journal}
  {\bibinfo  {journal} {SciPost Phys.}\ }\textbf {\bibinfo {volume} {7}},\
  \bibinfo {pages} {006} (\bibinfo {year} {2019})}\BibitemShut {NoStop}%
\bibitem [{\citenamefont {Hackenbroich}\ \emph {et~al.}(2018)\citenamefont
  {Hackenbroich}, \citenamefont {Sterdyniak},\ and\ \citenamefont
  {Schuch}}]{hackenbroich2018interplay}%
  \BibitemOpen
  \bibfield  {author} {\bibinfo {author} {\bibfnamefont {A.}~\bibnamefont
  {Hackenbroich}}, \bibinfo {author} {\bibfnamefont {A.}~\bibnamefont
  {Sterdyniak}},\ and\ \bibinfo {author} {\bibfnamefont {N.}~\bibnamefont
  {Schuch}},\ }\bibfield  {title} {\bibinfo {title} {Interplay of {SU}$(2)$,
  point group, and translational symmetry for projected entangled pair states:
  Application to a chiral spin liquid},\ }\href@noop {} {\bibfield  {journal}
  {\bibinfo  {journal} {Phys. Rev. B}\ }\textbf {\bibinfo {volume} {98}},\
  \bibinfo {pages} {085151} (\bibinfo {year} {2018})}\BibitemShut {NoStop}%
\bibitem [{\citenamefont {Li}\ and\ \citenamefont
  {Haldane}(2008)}]{li2008entanglement}%
  \BibitemOpen
  \bibfield  {author} {\bibinfo {author} {\bibfnamefont {H.}~\bibnamefont
  {Li}}\ and\ \bibinfo {author} {\bibfnamefont {F.~D.~M.}\ \bibnamefont
  {Haldane}},\ }\bibfield  {title} {\bibinfo {title} {Entanglement spectrum as
  a generalization of entanglement entropy: Identification of topological order
  in non-abelian fractional quantum hall effect states},\ }\href@noop {}
  {\bibfield  {journal} {\bibinfo  {journal} {Phys. Rev. Lett.}\ }\textbf
  {\bibinfo {volume} {101}},\ \bibinfo {pages} {010504} (\bibinfo {year}
  {2008})}\BibitemShut {NoStop}%
\bibitem [{\citenamefont {Cirac}\ \emph {et~al.}(2011)\citenamefont {Cirac},
  \citenamefont {Poilblanc}, \citenamefont {Schuch},\ and\ \citenamefont
  {Verstraete}}]{cirac2011entanglement}%
  \BibitemOpen
  \bibfield  {author} {\bibinfo {author} {\bibfnamefont {J.~I.}\ \bibnamefont
  {Cirac}}, \bibinfo {author} {\bibfnamefont {D.}~\bibnamefont {Poilblanc}},
  \bibinfo {author} {\bibfnamefont {N.}~\bibnamefont {Schuch}},\ and\ \bibinfo
  {author} {\bibfnamefont {F.}~\bibnamefont {Verstraete}},\ }\bibfield  {title}
  {\bibinfo {title} {Entanglement spectrum and boundary theories with projected
  entangled-pair states},\ }\href@noop {} {\bibfield  {journal} {\bibinfo
  {journal} {Phys. Rev. B}\ }\textbf {\bibinfo {volume} {83}},\ \bibinfo
  {pages} {245134} (\bibinfo {year} {2011})}\BibitemShut {NoStop}%
\bibitem [{\citenamefont {Poilblanc}\ \emph {et~al.}(2016)\citenamefont
  {Poilblanc}, \citenamefont {Schuch},\ and\ \citenamefont
  {Affleck}}]{poilblanc20162}%
  \BibitemOpen
  \bibfield  {author} {\bibinfo {author} {\bibfnamefont {D.}~\bibnamefont
  {Poilblanc}}, \bibinfo {author} {\bibfnamefont {N.}~\bibnamefont {Schuch}},\
  and\ \bibinfo {author} {\bibfnamefont {I.}~\bibnamefont {Affleck}},\
  }\bibfield  {title} {\bibinfo {title} {{SU}$(2)_1$ chiral edge modes of a
  critical spin liquid},\ }\href@noop {} {\bibfield  {journal} {\bibinfo
  {journal} {Phys. Rev. B}\ }\textbf {\bibinfo {volume} {93}},\ \bibinfo
  {pages} {174414} (\bibinfo {year} {2016})}\BibitemShut {NoStop}%
\bibitem [{Note1()}]{Note1}%
  \BibitemOpen
  \bibinfo {note} {Note the ES computed at $D=6$ does not show the CSL
  features, lacking well separated linear branches.}\BibitemShut {Stop}%
\bibitem [{Note2()}]{Note2}%
  \BibitemOpen
  \bibinfo {note} {In the supplemental materials, we provide more detailed
  results, including the finite-size analysis of the ES, CSL characteristics
  from the unrestricted simulations and more signatures of the phase
  transition.}\BibitemShut {Stop}%
\bibitem [{\citenamefont {Nishino}\ \emph {et~al.}(1996)\citenamefont
  {Nishino}, \citenamefont {Okunishi},\ and\ \citenamefont
  {Kikuchi}}]{nishino1996}%
  \BibitemOpen
  \bibfield  {author} {\bibinfo {author} {\bibfnamefont {T.}~\bibnamefont
  {Nishino}}, \bibinfo {author} {\bibfnamefont {K.}~\bibnamefont {Okunishi}},\
  and\ \bibinfo {author} {\bibfnamefont {M.}~\bibnamefont {Kikuchi}},\
  }\bibfield  {title} {\bibinfo {title} {Numerical renormalization group at
  criticality},\ }\href
  {https://doi.org/https://doi.org/10.1016/0375-9601(96)00128-4} {\bibfield
  {journal} {\bibinfo  {journal} {Physics Letters A}\ }\textbf {\bibinfo
  {volume} {213}},\ \bibinfo {pages} {69} (\bibinfo {year} {1996})}\BibitemShut
  {NoStop}%
\bibitem [{\citenamefont {He}\ and\ \citenamefont
  {Chen}(2015)}]{he2015distinct}%
  \BibitemOpen
  \bibfield  {author} {\bibinfo {author} {\bibfnamefont {Y.-C.}\ \bibnamefont
  {He}}\ and\ \bibinfo {author} {\bibfnamefont {Y.}~\bibnamefont {Chen}},\
  }\bibfield  {title} {\bibinfo {title} {Distinct spin liquids and their
  transitions in spin-1/2 {XXZ} kagome antiferromagnets},\ }\href@noop {}
  {\bibfield  {journal} {\bibinfo  {journal} {Phys. Rev. Lett.}\ }\textbf
  {\bibinfo {volume} {114}},\ \bibinfo {pages} {037201} (\bibinfo {year}
  {2015})}\BibitemShut {NoStop}%
\bibitem [{\citenamefont {He}\ \emph {et~al.}(2014)\citenamefont {He},
  \citenamefont {Sheng},\ and\ \citenamefont {Chen}}]{he2014chiral}%
  \BibitemOpen
  \bibfield  {author} {\bibinfo {author} {\bibfnamefont {Y.-C.}\ \bibnamefont
  {He}}, \bibinfo {author} {\bibfnamefont {D.}~\bibnamefont {Sheng}},\ and\
  \bibinfo {author} {\bibfnamefont {Y.}~\bibnamefont {Chen}},\ }\bibfield
  {title} {\bibinfo {title} {Chiral spin liquid in a frustrated anisotropic
  kagome {H}eisenberg model},\ }\href@noop {} {\bibfield  {journal} {\bibinfo
  {journal} {Phys. Rev. Lett.}\ }\textbf {\bibinfo {volume} {112}},\ \bibinfo
  {pages} {137202} (\bibinfo {year} {2014})}\BibitemShut {NoStop}%
\bibitem [{\citenamefont {Sun}\ \emph {et~al.}(2022)\citenamefont {Sun},
  \citenamefont {Jin}, \citenamefont {Tu},\ and\ \citenamefont
  {Zhou}}]{sun2022possible}%
  \BibitemOpen
  \bibfield  {author} {\bibinfo {author} {\bibfnamefont {R.-Y.}\ \bibnamefont
  {Sun}}, \bibinfo {author} {\bibfnamefont {H.-K.}\ \bibnamefont {Jin}},
  \bibinfo {author} {\bibfnamefont {H.-H.}\ \bibnamefont {Tu}},\ and\ \bibinfo
  {author} {\bibfnamefont {Y.}~\bibnamefont {Zhou}},\ }\bibfield  {title}
  {\bibinfo {title} {Possible chiral spin liquid state in the $ s= 1/2$ kagome
  {H}eisenberg model},\ }\href@noop {} {\bibfield  {journal} {\bibinfo
  {journal} {arXiv preprint arXiv:2203.07321}\ } (\bibinfo {year}
  {2022})}\BibitemShut {NoStop}%
\bibitem [{\citenamefont {Haghshenas}\ \emph {et~al.}(2019)\citenamefont
  {Haghshenas}, \citenamefont {Gong},\ and\ \citenamefont
  {Sheng}}]{haghshenas2019single}%
  \BibitemOpen
  \bibfield  {author} {\bibinfo {author} {\bibfnamefont {R.}~\bibnamefont
  {Haghshenas}}, \bibinfo {author} {\bibfnamefont {S.-S.}\ \bibnamefont
  {Gong}},\ and\ \bibinfo {author} {\bibfnamefont {D.}~\bibnamefont {Sheng}},\
  }\bibfield  {title} {\bibinfo {title} {Single-layer tensor network study of
  the {H}eisenberg model with chiral interactions on a kagome lattice},\
  }\href@noop {} {\bibfield  {journal} {\bibinfo  {journal} {Phys. Rev. B}\
  }\textbf {\bibinfo {volume} {99}},\ \bibinfo {pages} {174423} (\bibinfo
  {year} {2019})}\BibitemShut {NoStop}%
\bibitem [{\citenamefont {Weichselbaum}(2012)}]{Weichselbaum2012}%
  \BibitemOpen
  \bibfield  {author} {\bibinfo {author} {\bibfnamefont {A.}~\bibnamefont
  {Weichselbaum}},\ }\bibfield  {title} {\bibinfo {title} {Non-abelian
  symmetries in tensor networks: A quantum symmetry space approach},\ }\href
  {https://doi.org/https://doi.org/10.1016/j.aop.2012.07.009} {\bibfield
  {journal} {\bibinfo  {journal} {Ann. Phys.}\ }\textbf {\bibinfo {volume}
  {327}},\ \bibinfo {pages} {2972} (\bibinfo {year} {2012})}\BibitemShut
  {NoStop}%
\bibitem [{\citenamefont {Weichselbaum}(2020)}]{Weichselbaum2020}%
  \BibitemOpen
  \bibfield  {author} {\bibinfo {author} {\bibfnamefont {A.}~\bibnamefont
  {Weichselbaum}},\ }\bibfield  {title} {\bibinfo {title} {X-symbols for
  non-abelian symmetries in tensor networks},\ }\href
  {https://doi.org/10.1103/PhysRevResearch.2.023385} {\bibfield  {journal}
  {\bibinfo  {journal} {Phys. Rev. Research}\ }\textbf {\bibinfo {volume}
  {2}},\ \bibinfo {pages} {023385} (\bibinfo {year} {2020})}\BibitemShut
  {NoStop}%
\bibitem [{\citenamefont {Haegeman}()}]{tensorkit}%
  \BibitemOpen
  \bibfield  {author} {\bibinfo {author} {\bibfnamefont {J.}~\bibnamefont
  {Haegeman}},\ }\href {https://github.com/Jutho/TensorKit.jl} {\bibinfo
  {title} {{TensorKit.jl: A Julia package for large-scale tensor computations,
  with a hint of category theory.}}},\ \bibinfo {howpublished}
  {{https://github.com/Jutho/TensorKit.jl}}\BibitemShut {NoStop}%
\bibitem [{\citenamefont {Liao}\ \emph {et~al.}(2019)\citenamefont {Liao},
  \citenamefont {Liu}, \citenamefont {Wang},\ and\ \citenamefont
  {Xiang}}]{liao2019differentiable}%
  \BibitemOpen
  \bibfield  {author} {\bibinfo {author} {\bibfnamefont {H.-J.}\ \bibnamefont
  {Liao}}, \bibinfo {author} {\bibfnamefont {J.-G.}\ \bibnamefont {Liu}},
  \bibinfo {author} {\bibfnamefont {L.}~\bibnamefont {Wang}},\ and\ \bibinfo
  {author} {\bibfnamefont {T.}~\bibnamefont {Xiang}},\ }\bibfield  {title}
  {\bibinfo {title} {Differentiable programming tensor networks},\ }\href@noop
  {} {\bibfield  {journal} {\bibinfo  {journal} {Phys. Rev. X}\ }\textbf
  {\bibinfo {volume} {9}},\ \bibinfo {pages} {031041} (\bibinfo {year}
  {2019})}\BibitemShut {NoStop}%
\bibitem [{\citenamefont {Hasik}\ and\ \citenamefont
  {Mbeng}(2020)}]{hasik2020peps}%
  \BibitemOpen
  \bibfield  {author} {\bibinfo {author} {\bibfnamefont {J.}~\bibnamefont
  {Hasik}}\ and\ \bibinfo {author} {\bibfnamefont {G.}~\bibnamefont {Mbeng}},\
  }\href@noop {} {\bibinfo {title} {peps-torch: A differentiable tensor network
  library for two-dimensional lattice models}} (\bibinfo {year}
  {2020})\BibitemShut {NoStop}%
\end{thebibliography}%

\noindent

\renewcommand{\thesection}{S-\arabic{section}}
\setcounter{section}{0}  
\renewcommand{\theequation}{S\arabic{equation}}
\setcounter{equation}{0}  
\renewcommand{\thefigure}{S\arabic{figure}}
\setcounter{figure}{0}  

\indent

\section*{\Large\bf APPENDIX}
\subsection{CSL entanglement spectrum on finite-circumference cylinders}

\begin{figure}[!hbt]
  \includegraphics[width=\linewidth]{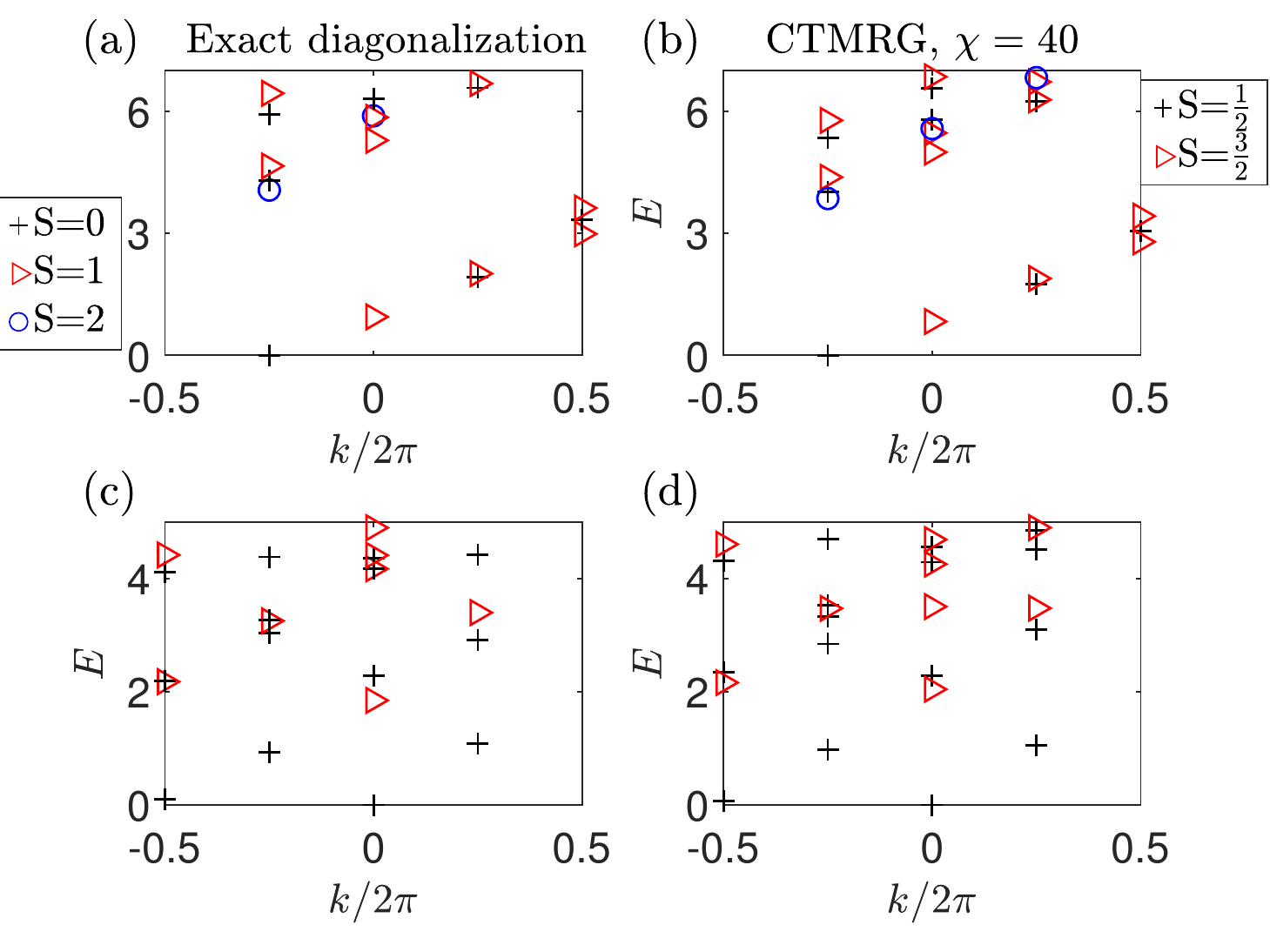}
  \caption{Entanglement spectrum of the optimized $\mathrm{A}_1+i\mathrm{A}_2$ chiral ansatz ($D=8$) at $\theta=0.2\pi$ computed on a $N_v=4$ infinitely-long cylinder. Left (a), (c): ED.  Right (b), (d): CTMRG with $\chi =40$. The even (first row) and odd (second row) sectors are normalized separately. }
\label{fig:Supp_fig1a}
\end{figure}

In the main text we used two methods, i.e. ED (exact contraction) and CTMRG, to obtain the boundary operators needed to calculate the ES on a (bi-partitioned) infinite cylinder of finite circumference $N_v$, see Fig.~2. The second method, which is approximate but applicable to larger $N_v$, is tested here by comparing its results to the one obtained by ED on the same thin cylinder ($N_v=4$). As shown in Fig.~\ref{fig:Supp_fig1a} (a)-(d) using the $D=8$ ansatz, a perfect agreement is found both for even and odd sectors in the low-energy regime. 

\begin{figure}[!hbt]
  \includegraphics[width=\linewidth]{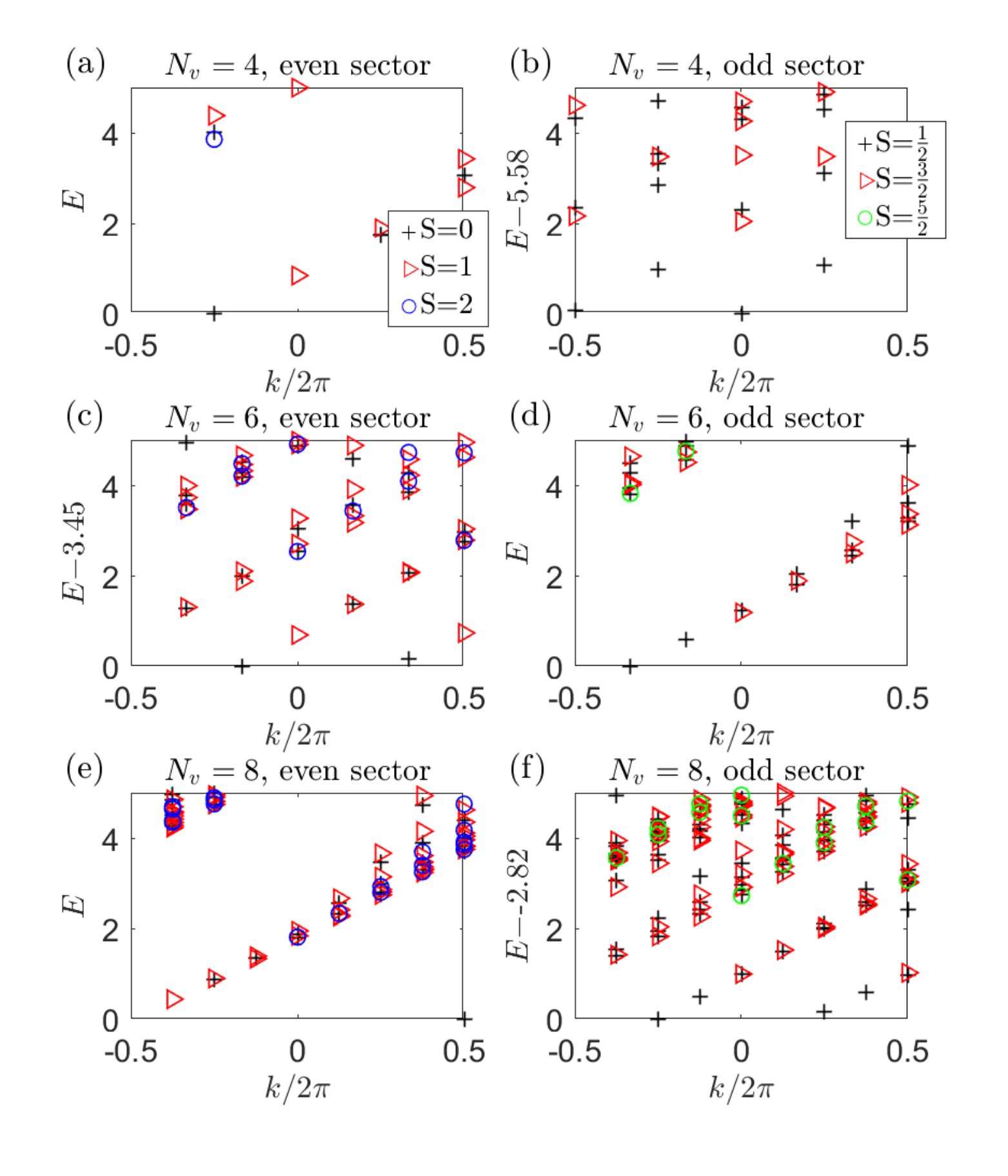}
  \caption{Entanglement spectrum of the optimized $\mathrm{A}_1+i\mathrm{A}_2$ chiral ansatz ($D=8$) at $\theta=0.2\pi$. CTMRG boundary tensors with $\chi =40$ are used. The even and odd sectors are normalized together. For $N_v=4m$ the leading sector is the even sector, while for $N_v=4m+2$ the leading sector is the odd sector, as reflected from the offsets on the energy axis introduced in (b), (c) and (f) (to move the ground state down to $E=0$).}
\label{fig:Supp_fig1b}
\end{figure}

The finite-circumference dependence of the two sectors of the ES can be seen in Fig.~\ref{fig:Supp_fig1b}. We observe a systematic even/odd effect depending on the parity of $N_v/2$: for $N_v=4m$ $(m\in \mathbb{N}^{+})$ the ES in the even sector has larger weight (lower energy) while for $N_v=4m+2$ the ES in the odd sector has larger weight (lower energy). We conjecture that this is because the leading sector in the two-site entanglement Hamiltonian is the odd sector. Moreover, we find the leading sector only has one dispersion branch while the sub-leading sector has two nearly-degenerate branches shifted by $\pi$ in momentum as can be seen in Fig.~\ref{fig:Supp_fig1b}. 

\subsection{Unconstrained iPESS simulation of the CSL state}

\begin{figure}[t]
  \includegraphics[width=\linewidth]{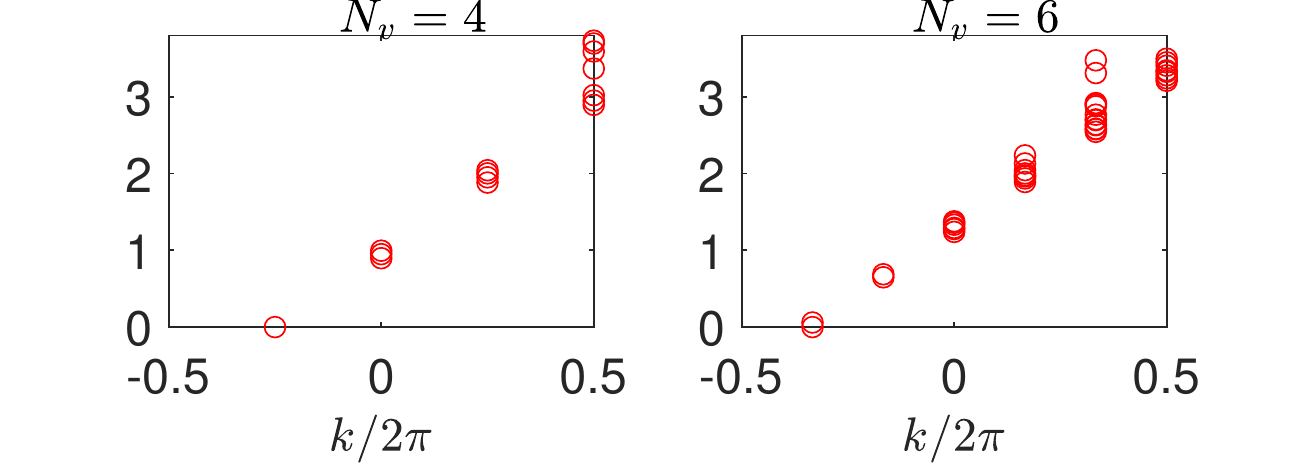}
  \caption{Entanglement spectrum of the optimized unconstrained $D=8$ iPESS at $\theta=0.2\pi$. The fixed point boundary operators are obtained from CTMRG $\chi =40$.}\label{fig:Supp_fig2}
\end{figure}

Here we show that the kagome CSL state can also be obtained from simulations without symmetry constraints, similar to the case on the square lattice~\cite{hasik2022}. In this calculation, we still consider the same parameter $\theta=0.2\pi$, and perform gradient optimization with the gradient obtained from an automatic difference algorithm~\cite{liao2019differentiable,hasik2020peps}.  The $D=8$ state is optimized at $\chi=64$ and further evaluated at $\chi=128$. The energy per site $E=-0.5018$ is only slightly better than the energy of the symmetric ansatz, via forming a small magnetization around $5\times 10^{-3}$. This feature of CSL phase is in sharp contrast to the kagome spin liquid at the KHA point where the finite $D$ energy can be improved a lot through forming considerable mean-field like magnetization~\cite{KHA2017gap,KHA2017gapless}. The ES for the unconstrained simulation is shown in Fig.~\ref{fig:Supp_fig2}. As the SU$(2)$ symmetry is slightly broken, the spin quantum number is not well-defined and we can not distinguish topological sectors. Nevertheless, the dispersion is perfectly linear and the degeneracies of low energy states agree well with those in the leading sectors in Fig.~\ref{fig:Supp_fig1b}. We also find that the correlation functions have the same long-range features as the symmetric states in the main text (not shown here).

\subsection{Behavior of the variational parameters across the phase transition}
\begin{figure*}
  \includegraphics[width=2\columnwidth]{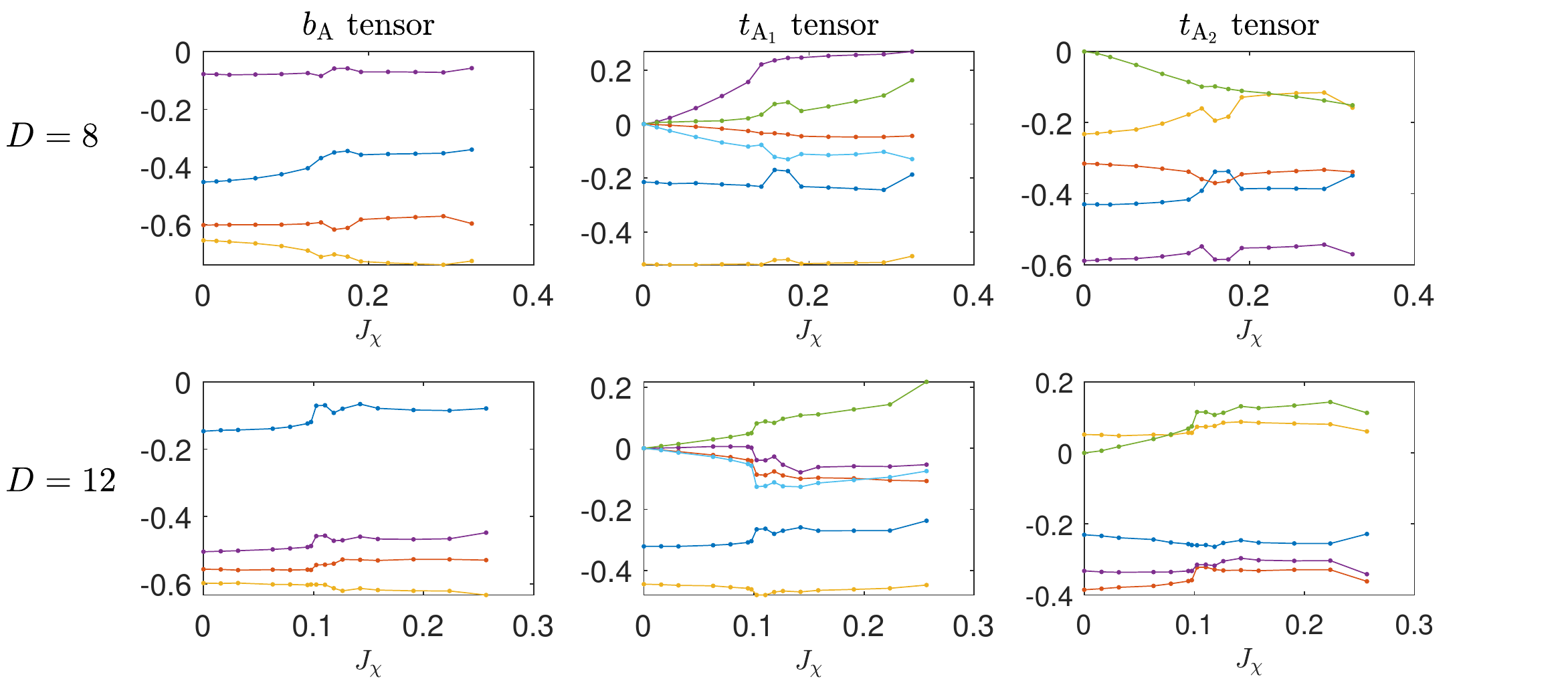}
  \caption{Optimized variational parameters associated to the $b_{\mathrm{A}},\,t_{\mathrm{A}_1}$ and $t_{\mathrm{A}_2}$ tensors versus $J_{\chi}$. At the $J_{\chi}=0$ point, the parameters defining the  $t_{\mathrm{A}_1}^{\mathrm{II}},t_{\mathrm{A}_2}^{\mathrm{I}}$ tensors vanish exactly (the optimal ansatz there is the reflection symmetric ansatz $t_{\mathrm{A}_1}^{\mathrm{I}}+t_{\mathrm{A}_2}^{\mathrm{II}}$). When $J_{\chi}$ is turned on, these parameters grow linearly up to the critical point.}\label{fig:variational_parameters}
\end{figure*}

\begin{figure*}[!h]
  \includegraphics[width=\linewidth]{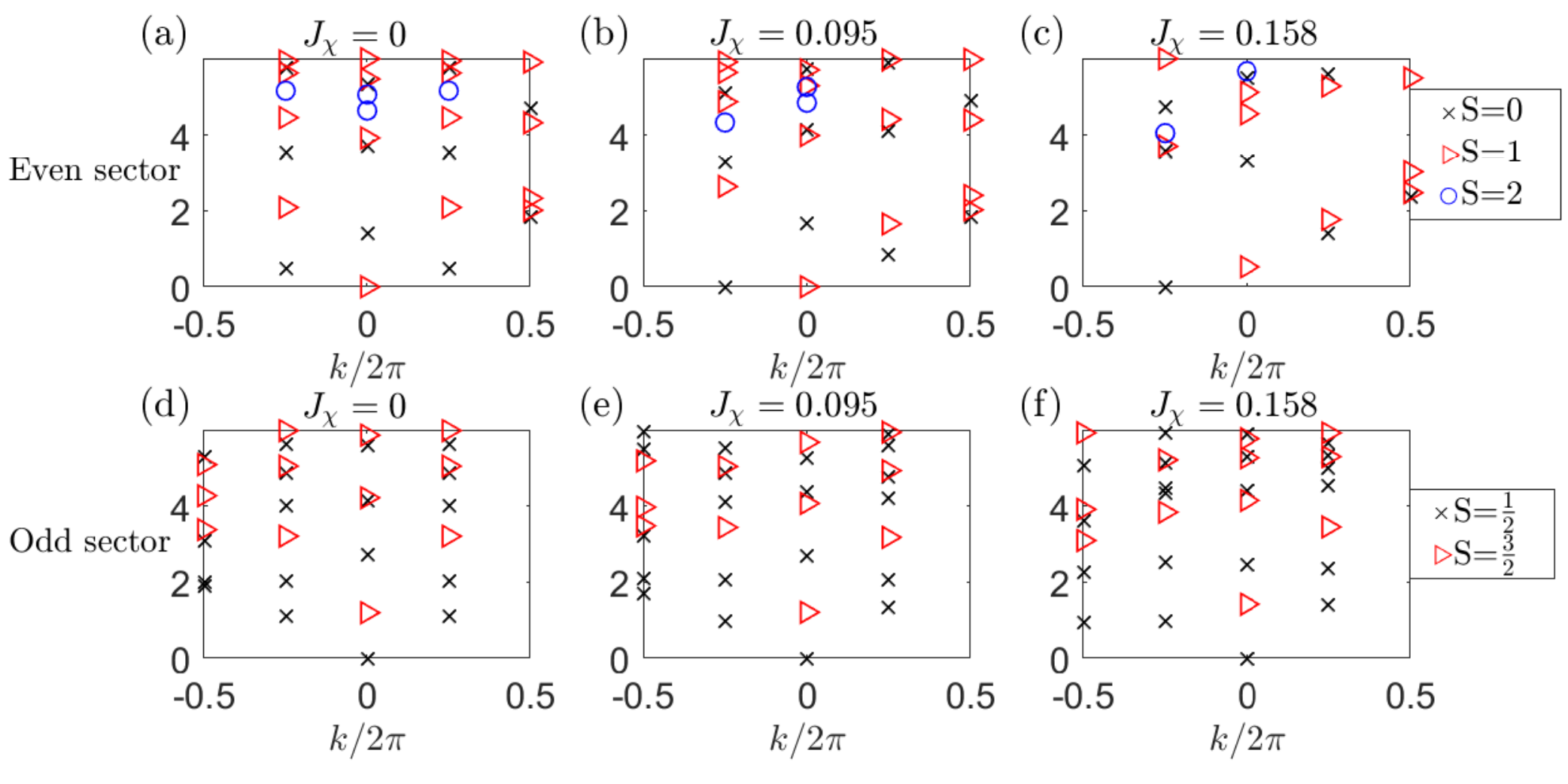}
  \caption{Entanglement spectrum of the optimized $\mathrm{A}_1+i\mathrm{A}_2$ chiral ansatz ($D=8$) at different $J_{\chi}$. $N_v=4$ and CTMRG $\chi =40$ are used. First column (a), (d): ES at the KHA point with $J_{\chi}=0$. Second column (b), (e): ES below the $D=8$ critical $J_\chi$ point. Third column (c)-(f): ES above the $D=8$ critical $J_\chi$ point.}\label{fig:Supp_fig3}
\end{figure*}

\begin{figure*}[!htb]
  \includegraphics[width=\linewidth]{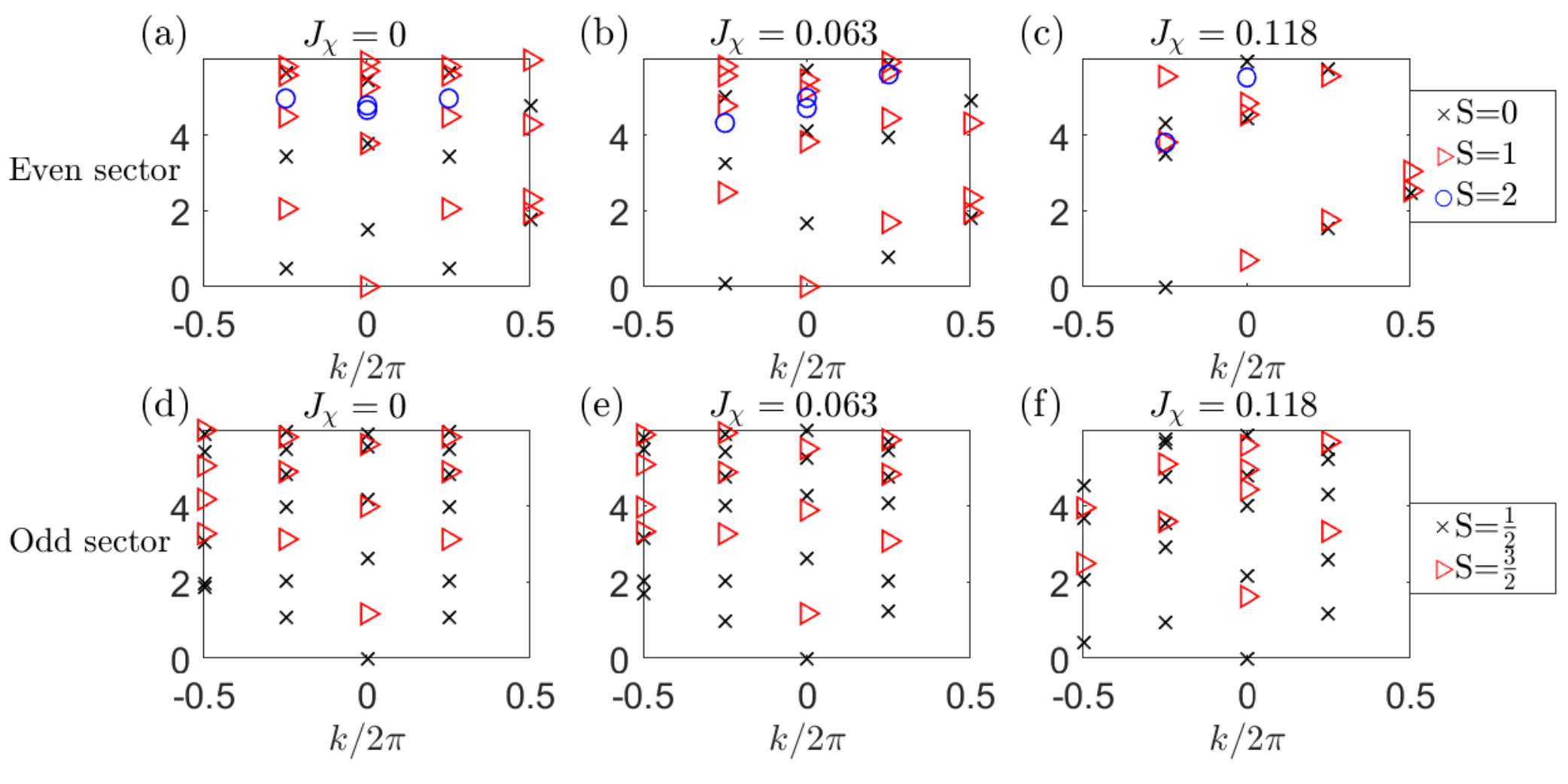}
  \caption{Entanglement spectrum of the optimized $\mathrm{A}_1+i\mathrm{A}_2$ chiral ansatz ($D=12$) at different $J_{\chi}$. $N_v=4$ and CTMRG $\chi =40$ are used. First column (a), (d): ES at the KHA point with $J_{\chi}=0$. Second column (b), (e): ES below the $D=12$ critical $J_\chi$ point. Third column (c)-(f): ES above the $D=12$ critical $J_\chi$ point.}\label{fig:Supp_fig4}
\end{figure*}

Apart from the chirality observable $\langle\chi_{ijk}\rangle$ displayed in the main text, the phase transition from tuning $J_{\chi}$ is also reflected in the behavior of the variational parameters defining the tensors, as can be seen in Fig.~\ref{fig:variational_parameters}. Before the transition, the parameters vary smoothly and the parameters associated to the  $t_{\mathrm{A}_1}^{\mathrm{II}},t_{\mathrm{A}_2}^{\mathrm{I}}$ tensors grow from zero linearly. At the transition point, some of the parameters show clear jumps (e.g., the parameters in $t_{\mathrm{A}_1}^{\mathrm{II}}$ tensors at $D=12$). Below and above the transition, the parameters have two distinct behaviors, which is consistent with the level crossing picture in first-order phase transition.

\subsection{Behavior of the entanglement spectrum across the phase transition}

The two phases can be further distinguished from their ES. First, we show the ES at the KHA point $J_{\chi}=0$ for $D=8,12$ in Fig.~\ref{fig:Supp_fig3} (a)-(b) and Fig.~\ref{fig:Supp_fig4} (a)-(b). The quantum numbers and dispersions are essentially the same as the $D=3$ short-range RVB state given in Ref.~\cite{poilblanc2012topological}, as can be expected from the fact that the SU$(2)$ symmetric variational state at finite (small) $D$ has $\mathbb{Z}_2$ topological order~\cite{KHA2017gap}. We further show the ES for $J_{\chi}$ values below and above the (estimated) critical value of the transition for $D=8$ in Fig.~\ref{fig:Supp_fig3} (c)-(f) and for $D=12$ in Fig.~\ref{fig:Supp_fig4} (c)-(f). It is clear that the ES in the leading sector (even sector for $N_v=4$ here) show a qualitative change compared to the non-chiral case. Once $J_{\chi}$ is turned on, while still below the transition, there is no separation between low and high energy states and the lowest entanglement energy level is a $S=1$ triplet (although reflection symmetry is broken). Above the transition, the low energy chiral branches become well separated from the higher energy states with the lowest level being a $S=0$ singlet. For the sub-leading sector (odd sector for $N_v=4$ here) which has smaller weight, as there exists two branches, the transition is not so clear. We expect the transition in the sub-leading sector will become  more visible for larger cylinder circumferences $N_v$.

\end{document}